\documentclass[preprintnumbers,
amsmath,
prd,nofootinbib,floatfix,11pt,
]{revtex4}

\newcommand\beq{\begin{eqnarray}}
\newcommand\eeq{\end{eqnarray}}

\def\muPhi{{\mu_\Phi}}
\def\bphi{{b_\phi}}

\def\lsim{\mathrel{\rlap{\lower4pt\hbox{$\sim$}}
    \raise1pt\hbox{$<$}}}                
\def\gsim{\mathrel{\rlap{\lower4pt\hbox{$\sim$}}
    \raise1pt\hbox{$>$}}}            

\allowdisplaybreaks
\usepackage{graphicx}
\usepackage{setspace}

\begin{document}
\renewcommand{\theequation}{\arabic{section}.\arabic{equation}}
\renewcommand{\thefigure}{\arabic{section}.\arabic{figure}}
\renewcommand{\thetable}{\arabic{section}.\arabic{table}}

\title{\Large \baselineskip=16pt 
\mbox{Mixed gluinos and sgluons from a new $SU(3)$ gauge group}}

\author{Stephen P.~Martin}
\affiliation{
\mbox{\it Department of Physics, Northern Illinois University, DeKalb IL 60115}}

\begin{abstract}\normalsize \baselineskip=15pt 
I study supersymmetric models in which the QCD gauge group is the remnant diagonal 
subgroup from the spontaneous breaking of an $SU(3) \times SU(3)$ gauge group at a multi-TeV scale. 
In renormalizable models with soft supersymmetry breaking, the scalar potential is shown to have
global minima with the required gauge symmetry breaking pattern. In addition to a massive 
color octet vector boson, this framework predicts 3 color octet spin-0 sgluons, and 4 color octet gluinos
with both Dirac and Majorana mass terms. One of the gluino mass eigenstates typically has a coupling to quark-squark pairs that is at least as large as the prediction of minimal supersymmetry, 
but it need not be the lightest one.
\end{abstract}

\maketitle
\tableofcontents

\baselineskip=15.4pt

\newpage

\section{Introduction\label{sec:intro}}
\setcounter{equation}{0}
\setcounter{figure}{0}
\setcounter{table}{0}
\setcounter{footnote}{1}

The Large Hadron Collider (LHC) has discovered the Higgs boson 
associated with electroweak symmetry breaking,
but so far has not provided any insight into the hierarchy problem. Instead, it has imposed significant
lower limits on the masses of supersymmetric particles, and more generally 
on any new particles that could be involved in
symmetries or dynamics that might explain why the electroweak scale is so much lighter 
than the Planck scale and other high mass scales associated with new physics. 
At the same time, a mass near 125 GeV for the lightest Higgs boson,
within the context of supersymmetric extensions of the Standard model, suggests that the top squark
and other superpartner masses could very well lie at a characteristic scale $M_{\rm SUSY}$ in the multi-TeV range, beyond the reach of the 14 TeV LHC.

One possibility is that supersymmetry really is the essential part of 
the explanation for the big hierarchy problem 
$M_Z^2 \ll M_{\rm Planck}$, but that there is 
some other subsidiary principle,\footnote{It is also possible that the
little hierarchy is just the result of a coincidence. This should be taken seriously because it is 
enormously less severe than the big hierarchy problem. However, there is no objective, 
scientific way of deciding how much of a coincidence is too severe; in my view this is a personal 
and inherently subjective choice 
that scientists must nevertheless make in order to decide how to allocate 
limited resources such as time and money.} 
not yet understood, that could explain the 
little hierarchy problem $M_Z^2 \ll M_{\rm SUSY}^2$. 
A common feature to be expected in that case 
is that the minimal supersymmetric standard model (MSSM) should be extended beyond the 
minimal particle content
in the multi-TeV mass range or below. Attempts along these lines are far too numerous to review here.

The MSSM already contains one vectorlike combination of fields, the Higgs supermultiplets $H_u$ and $H_d$,
which obtain a bare supersymmetry-preserving mass term, $\mu$. Whatever mechanism is responsible for
ensuring that $\mu$ is non-zero but also not far above the TeV scale could plausibly 
also be responsible for placing other vectorlike chiral supermultiplets at the TeV scale. 
In the same spirit, one can also suppose that there are other gauge supermultiplets that 
obtain masses at the TeV scale, where the corresponding gauge symmetries are spontaneously broken.
In this paper, I will consider one such possibility that has already been widely
considered 
\cite{Pati:1975ze,Preskill:1980mz,Hall:1985wz,Frampton:1987dn,Frampton:1987ut,Bagger:1987fz,Hill:1991at,Martin:1992aq,Hill:1993hs,
Dobrescu:1997nm,Chivukula:1998wd,
Chivukula:1996yr,Simmons:1996fz,Bai:2010dj,Chivukula:2013xka,Chivukula:2014rka,
Chen:2014haa,Chivukula:2015kua,
Bai:2017zhj,Agrawal:2017ksf,Bai:2018jsr,Bai:2018wnt}
in the non-supersymmetric context: that the QCD $SU(3)_C$ gauge group of the MSSM and the
Standard Model is the remnant of a spontaneous breaking of the type:
\beq
SU(3)_A \times SU(3)_B &\rightarrow& SU(3)_C .
\eeq
The gauge bosons associated with the 
diagonal subgroup of $SU(3)_A \times SU(3)_B$ are the massless gluons of the Standard Model. 
The remaining 8 vector bosons have been variously referred to in the literature as 
axigluons \cite{Pati:1975ze,Preskill:1980mz,Hall:1985wz,Frampton:1987dn,Frampton:1987ut,Bagger:1987fz}
or topgluons \cite{Hill:1991at,Martin:1992aq,Hill:1993hs,
Dobrescu:1997nm,Chivukula:1998wd}
or colorons 
\cite{Chivukula:1996yr,Simmons:1996fz,Bai:2010dj,Chivukula:2013xka,Chivukula:2014rka,
Chen:2014haa,Chivukula:2015kua,Bai:2017zhj,Agrawal:2017ksf,Bai:2018jsr,Bai:2018wnt}, 
depending on how the
Standard Model fermions are assigned to $SU(3)_A$ and $SU(3)_B$ representations. Here, I 
will study the possibility of realizing this symmetry breaking consistently in a renormalizable 
softly broken supersymmetric model. This requires the presence of two chiral 
superfields that transform as the
fundamental and anti-fundamental representations of both gauge groups, to be denoted in this paper as:
\beq
\Phi_j^k \sim ({\bf 3},\> {\bf \overline 3})
,
\qquad\qquad
\overline\Phi_k^j \sim ({\bf \overline 3},\> {\bf 3})
.
\label{eq:definePhiPhibar}
\eeq
A lowered index corresponds to a fundamental ${\bf 3}$ representation of $SU(3)$,
and a raised index to an anti-fundamental ${\bf \overline 3}$.
Thus, in both instances in eq.~(\ref{eq:definePhiPhibar}), $j$ is an $SU(3)_A$ index, and $k$ is an $SU(3)_B$ index. 

This supersymmetric model then predicts the existence of, in addition the coloron
vector boson $X$, four gluino mass eigenstates (including an admixture of what can be regarded
as the MSSM gluino) with both Dirac and Majorana mass contributions, three color octet scalars (sgluons), two color singlet fermions (singlinos)
and four real scalar singlets, in addition to the usual superpartners of the MSSM. Models with 
Dirac mass terms for gauginos 
have a long history 
\cite{Fayet:1978qc,Polchinski:1982an,Hall:1990hq,Jack:1999ud,Jack:1999fa,
supersoft,Chacko:2004mi,Antoniadis:2005em,Kribs:2007ac,
Amigo:2008rc,Benakli:2008pg,Carpenter:2010as,Kribs:2010md,Abel:2011dc,Fok:2012fb,
Csaki:2013fla,Dudas:2013gga,Benakli:2014cia,Nelson:2015cea,Alves:2015kia,
Martin:2015eca,Chakraborty:2018izc}.
The present paper is an alternative
to models where Dirac gaugino masses arise due to 
supersoft \cite{supersoft} supersymmetry breaking following from $D$-term breaking 
and feature a continuous $R$ symmetry \cite{Hall:1990hq,supersoft,Kribs:2007ac}, 
where the gauge supermultiplet sector can be considered as $N=2$ supersymmetry multiplets.
Instead, the Dirac gluino mass parameters here arise from an additional gauge group and 
the chiral fermions associated with its breaking. There are also Majorana gluino masses,
so that the gluinos are mixed.

I now discuss some other conventions and notations to be used below. Adjoint representation indices of
$SU(3)$ are represented by letters $a,b,c,\ldots$.
The generators of the fundamental representation are $T_j^{ak}$, and obey the general
$SU(N_c)$ trace, commutator, anti-commutator, and Fierz identities:
\beq
{\rm Tr}[T^a T^b] &=& \frac{1}{2} \delta^{ab}
,
\label{eq:Dynkinindex}
\\
{}
[T^a, T^b] _j{}^k &=& i f^{abc} T_j^{ck}
,
\label{eq:deffabc}
\\
\{T^a, T^b\}_j{}^k &=& \frac{1}{N_c} \delta^{ab} \delta_j^k + d^{abc} T_j^{ck}
,
\label{eq:defdabc}
\\
T_j^{ak} T_l^{am} &=& \frac{1}{2} \delta_j^m \delta_l^k - \frac{1}{2N_c} \delta_j^k \delta_l^m
.
\eeq
Here eq.~(\ref{eq:Dynkinindex}) establishes the usual normalization of the generators,
while eq.~(\ref{eq:deffabc}) defines the anti-symmetric structure constants $f^{abc}$ and eq.~(\ref{eq:defdabc}) defines the symmetric anomaly coefficients $d^{abc}$.
There follows:
\beq
{\rm Tr}[T^a T^b T^c] &=& \frac{1}{4} \left (d^{abc} + i f^{abc} \right )
,
\\
T_j^{ak} T_k^{al} &=& \frac{N_c^2 - 1}{2N_c} \delta_j^l  
.
\eeq
Also, for $N_c=3$ only (as assumed from now on), 
there are the anti-symmetric tensor invariant symbols $\epsilon^{jkl}$ 
and $\epsilon_{jkl}$, which by convention are taken here to have
\beq
\epsilon^{123} = \epsilon_{123} = 1.
\eeq
Then one also has the useful identity:
\beq
\epsilon_{jlm} \epsilon^{knp} T_n^{al} T_p^{bm} &=&
d^{abc} T_j^{ck} - \frac{1}{6} \delta^{ab} \delta_j^k
.
\eeq 
The notations and conventions for supersymmetry and 2-component fermions follow those in 
\cite{primer}.

For an appropriate choice of potential parameters, as demonstrated below,
the scalar components of $\Phi$ and $\overline \Phi$ will 
acquire vacuum expectation values (VEVs) of the form
\beq
\langle \phi_j^k \rangle = \delta_j^k v ,
\qquad\qquad
\langle \overline \phi_j^k \rangle = \delta_j^k \overline v.
\eeq
In that case, the massless gluon field $G$ and the massive color octet vector field $X$ are related
to the $SU(3)_A$ and $SU(3)_B$ gauge vector fields by:
\beq
\begin{pmatrix}
G_\mu^a
\\
X_\mu^a 
\end{pmatrix}
&=&
\begin{pmatrix}
\cos\theta & \phantom{x}\sin\theta
\\
-\sin\theta & \phantom{x}\cos\theta
\end{pmatrix}
\begin{pmatrix}
A_\mu^a
\\
B_\mu^a 
\end{pmatrix}
,
\eeq
where
\beq
\sin\theta &=& g_A/\sqrt{g_A^2 + g_B^2}
,
\qquad\quad
\cos\theta \>=\> g_B/\sqrt{g_A^2 + g_B^2},
\eeq
and $X$ obtains a squared mass: 
\beq
M_X^2 = (g_A^2 + g_B^2) (|v|^2 + |\overline v|^2) .
\label{eq:M2vector}
\eeq
The QCD coupling is related to the original gauge couplings by
\beq
g_3 &=& {g_A g_B}/{\sqrt{g_A^2 + g_B^2}},
\label{eq:gCfromgAgB}
\eeq
and fields that transform as $(R_A, R_B)$ under $SU(3)_A \times SU(3)_B$ will transform as
the (reducible, in general) representation $R_A \times R_B$ of $SU(3)_C$. In particular,
for quarks originally in the fundamental ${\bf 3}$ representation of the 
$SU(3)_A$ gauge group, the covariant derivative is:
\beq
D_\mu q_j = (\partial_\mu q_j - i g_3 G^a_\mu T_j^{ak} q_k)
+ i g_3 \tan\theta\> X^a_\mu T_j^{ak} q_k
.
\eeq
On the other hand, for quarks originally in the fundamental representation of $SU(3)_B$,
then we have
\beq
D_\mu q_j = (\partial_\mu q_j - i g_3 G^a_\mu T_j^{ak} q_k)
- i g_3 \cot\theta\> X^a_\mu T_j^{ak} q_k
.
\eeq
In the following, I will assume that all of the Standard Model quarks and their superpartners
live in the fundamental representation of $SU(3)_A$, although this is not inevitable. There can also be 
additional vectorlike quarks and squarks transforming under $SU(3)_B$, and these will indeed play a role in
section \ref{sec:realmodel}. These can mix with the usual quarks 
by Yukawa couplings to the $\Phi$ and $\overline \Phi$ fields, 
breaking flavor symmetries and thus allowing 
the vectorlike quarks to decay. For simplicity, it is assumed that these Yukawa couplings are 
non-zero but very small, as is technically natural. 

In the remainder of this paper, I will explore one possible supersymmetric 
setup that is renormalizable and avoids fundamental singlets (with potentially dangerous tadpoles). 
I will show that the symmetry breaking pattern given above can 
indeed be realized in a stable vacuum that is the global minimum of the potential. 
In similar non-supersymmetric models,
the minimization of the potential has been analyzed in ref.~\cite{Bai:2017zhj}. 
However, the softly broken 
supersymmetric case is quite different
because two Higgsing fields $\Phi, \overline \Phi$ are required by the anomaly cancellation
associated with the fermionic components, and because 
the structure of the dimensionless couplings in the scalar potential is constrained as dictated
by supersymmetry.
The resulting theory
naturally includes Dirac masses for the MSSM gluinos along with the usual Majorana masses. The lightest of the mixed gluino states can be significantly 
lighter than the color octet vectors $X$ and the spin-0 sgluons. There are also inevitably new
color singlet scalars and fermions. The 
phenomenology of these states will be briefly considered in section \ref{sec:pheno}.

\section{Supersymmetric models with $SU(3)_A \times SU(3)_B \rightarrow SU(3)_C$\label{sec:model}}
\setcounter{equation}{0}
\setcounter{figure}{0}
\setcounter{table}{0}
\setcounter{footnote}{1}

Consider a model consisting of the MSSM and the fields $\Phi$ and $\overline \Phi$.
The most general renormalizable superpotential of this theory is:
\beq
W &=& 
\frac{1}{6} \epsilon^{jkl} \epsilon_{mnp} \left (
y\, \Phi_j^m \Phi_k^n \Phi_l^p 
+
\overline y \, \overline\Phi_j^m \overline\Phi_k^n \overline\Phi_l^p \right )
-\muPhi \Phi_j^k \overline \Phi_k^j 
+ W_{\rm MSSM},
\eeq
where $y$ and $\overline y$ are Yukawa couplings and $\mu_\Phi$ is a mass term, which 
is analogous to the $\mu$ term of the MSSM, and can be presumed
to have the same sort of origin. As a very rough estimate, $\mu_\Phi$
can therefore be taken to be of order a multi-TeV scale. 
The existence of $y$ and $\overline y$
relies on the fact that the gauge groups are $SU(3)$, because only in this case among the special
unitary groups
does the invariant
symbol $\epsilon^{jkl}$ exist, corresponding to the group theory fact that the antisymmetric
product of fundamental representations 
${\bf 3} \times {\bf 3} \times {\bf 3}$ contains a singlet.  
As a convention, $y$ and $\overline y$ can be taken real and positive 
without loss of generality; then the phase of $\mu_\Phi$ is physical.
The soft supersymmetry breaking Lagrangian is:
\beq
{\cal L}_{\rm soft} &=& 
\biggl [ 
-\frac{1}{2} M_A \lambda^a_A \lambda^a_A
-\frac{1}{2} M_B \lambda^a_B \lambda^a_B
- \frac{1}{6} \epsilon^{jkl} \epsilon_{mnp}
\left (a \, \phi_j^m \phi_k^n \phi_l^p 
+ \overline a \,\overline\phi_j^m \overline\phi_k^n \overline\phi_l^p \right )
+ \bphi \phi_j^k \overline \phi_k^j 
\biggr ] + {\rm c.c.}
\nonumber 
\\
&&
- m^2 (\phi_j^k)^* \phi_j^k
- \overline m^2 (\overline \phi_j^k)^* \overline \phi_j^k
,
\eeq
where $\lambda_A^a$ and $\lambda_B^a$ are the gauginos for the $SU(3)_A$ and $SU(3)_B$ gauge groups
respectively.

One can now expand the scalar fields around diagonal VEVs:
\beq
\phi_j^k &=&  
\delta_j^k \left ( 
v + \frac{\phi_0}{\sqrt{3}}
\right )
+ \sqrt{2}\, T_j^{ak} \phi^a
,
\\
\overline\phi_j^k &=&
\delta_j^k \left ( 
\overline{v} + \frac{\overline{\phi}_0}{\sqrt{3}}
\right )
+ \sqrt{2}\, T_j^{ak} \overline{\phi}^a
,
\eeq
where $\phi_0$, $\overline\phi_0$ and $\phi^a$, $\overline\phi^a$ are 
complex scalar fields with canonically normalized
kinetic terms, which live in the singlet and adjoint representations of $SU(3)_C$.
Similarly, the fermionic components of $\Phi$ and $\overline \Phi$ can be expanded as:
\beq
\psi_j^k &=&  
\frac{1}{\sqrt{3}} \,\delta_j^k\, \psi_0
+ \sqrt{2} \,
T_j^{ak} \psi^a
,
\\
\overline\psi_j^k &=&
\frac{1}{\sqrt{3}} \,\delta_j^k\, \overline{\psi}_0
+ \sqrt{2} \,
T_j^{ak} \overline{\psi}^a
,
\eeq
where $\psi_0$ and $\psi^a$ and $\overline\psi_0$ and $\overline\psi^a$ 
are 2-component fermion fields with canonically normalized 
kinetic terms.

The interactions of the new fermions with the scalars and their VEVs are 
\beq
{\cal L} &=&
\biggl [
(\overline v + \overline\phi_0/\sqrt{3})^* 
(g_A \lambda_A^a \overline\psi^a - g_B \lambda_B^a \overline\psi^a)
+ (v + \phi_0/\sqrt{3})^* (g_B \lambda_B^a \psi^a - g_A \lambda_A^a \psi^a) 
\nonumber \\ &&
+ \frac{1}{\sqrt{2}} (d^{abc} + i f^{abc}) \left (
g_A \overline\phi^{a*} \overline\psi^b \lambda_A^c
-g_A \phi^{a*} \lambda_A^b \psi^c 
+ g_B \phi^{a*} \psi^b \lambda_B^c
- g_B \overline\phi^{a*} \lambda_B^b \overline\psi^c
\right )
\nonumber \\ &&
+ \frac{g_A}{\sqrt{3}} \left (
\overline \phi^{a*} \lambda_A^a \overline\psi_0
- \phi^{a*} \lambda_A^a \psi_0
\right )
+ \frac{g_B}{\sqrt{3}} \left (
\phi^{a*} \lambda_B^a \psi_0 -\overline \phi^{a*} \lambda_B^a \overline\psi_0
\right )
\biggr ] + {\rm c.c.}
\label{eq:LSFFgauge}
\eeq
from gaugino-fermion-scalar interactions, and
\beq
{\cal L} &=&
\biggl \{
y \left [
-(v + \phi_0/\sqrt{3}) \psi_0 \psi_0 
+ \frac{1}{2} (v + \phi_0/\sqrt{3}) \psi^a \psi^a
+ \frac{1}{\sqrt{3}} \phi^a \psi_0 \psi^a
- \frac{1}{\sqrt{2}} d^{abc} \phi^a \psi^b \psi^c \right ]
\nonumber \\ &&
+ 
\overline y \left [
-(\overline v + \overline \phi_0/\sqrt{3}) \overline\psi_0 \overline\psi_0 
+ \frac{1}{2} (\overline v + \overline \phi_0/\sqrt{3}) \overline\psi^a \overline\psi^a
+ \frac{1}{\sqrt{3}} \overline\phi^a \overline\psi_0 \overline\psi^a
- \frac{1}{\sqrt{2}} d^{abc} \overline\phi^a \overline\psi^b \overline\psi^c \right ]
\nonumber \\ &&
+ \mu_\Phi \bigl [\psi_0 \overline \psi_0
+ \psi^a \overline \psi^a \bigr ]
\biggr \} + {\rm c.c.}
\label{eq:LSFFyukawa}
\eeq
from the superpotential.
The mass eigenstates are then obtained as follows. There are four 2-component $SU(3)_C$-octet
fermions (gluinos), with mass matrix
in the basis $\tilde g^a = (\lambda_A^a,\> \lambda_B^a,\>  \psi^a,\>  \overline\psi^a)$:
\beq
M_{\tilde g} &=& 
\begin{pmatrix}
M_A & \phantom{-}0 & \phantom{-}g_A v^*\phantom{.} & -g_A \overline v^*\phantom{.}
\\
0 & \phantom{-}M_B & -g_B v^* & \phantom{-}g_B \overline v^*
\\
\phantom{-}g_A v^* & -g_B v^* & -y v & -\muPhi
\\
-g_A \overline v^* & \phantom{-}g_B \overline v^* & -\muPhi & -\overline y\hspace{1pt} \overline v
\end{pmatrix} .
\eeq
This can be diagonalized by a unitary matrix $U$ to obtain the mass eigenvalues:
\beq
M_{\tilde g}^{\rm diag} = U^* M_{\tilde g} U^\dagger .
\label{eq:defineUmatrix}
\eeq
There are also two gauge-singlet 2-component fermions, which in the basis 
$\tilde \chi = (\psi_0,\>  \overline\psi_0)$ have a mass matrix
\beq
M_{\tilde \chi} &=& 
\begin{pmatrix}
2 y v & -\muPhi
\\
-\muPhi & 2 \overline y\hspace{1pt} \overline v  
\end{pmatrix},
\eeq
with squared mass eigenvalues
\beq
|\mu_\Phi|^2 + 2 |yv|^2 + 2 |\overline y \overline v|^2
\pm 2 \sqrt{
|y v \mu_\Phi^* + \overline y^* \overline v^* \mu_\Phi |^2
+ (|yv|^2 - |\overline y \overline v|^2)^2
}.
\eeq

In order to obtain the new scalar mass eigenvalues, one can proceed by first obtaining the scalar potential
\beq
V = V_D + V_F + V_{\rm soft}
\label{eq:Vtotal}
\eeq
as a function of the canonically normalized fields. The supersymmetric $D$-term contribution
is 
\beq
V_D &=& \frac{1}{2} (D_A^a D_A^a + D_B^a D_B^a),
\eeq
where
\beq
D_A^a &=& \frac{g_A}{\sqrt{2}} 
\Bigl [
(\overline v + \overline\phi_0/\sqrt{3}) \overline\phi^{a*}
+ (\overline v + \overline\phi_0/\sqrt{3})^* \overline\phi^{a}
- (v + \phi_0/\sqrt{3}) \phi^{a*}
- (v + \phi_0/\sqrt{3})^* \phi^{a}
\nonumber \\ &&
+ \frac{1}{\sqrt{2}} \left (d^{abc} + i f^{abc} \right )
\left (\overline\phi^{b*} \overline\phi^c - \phi^b \phi^{c*} \right )
\Bigr ]
,
\\
D_B^a &=& \frac{g_B}{\sqrt{2}} 
\Bigl [
(v + \phi_0/\sqrt{3}) \phi^{a*}
+ (v + \phi_0/\sqrt{3})^* \phi^{a}
- (\overline v + \overline\phi_0/\sqrt{3}) \overline\phi^{a*}
- (\overline v + \overline\phi_0/\sqrt{3})^* \overline\phi^{a}
\nonumber \\ &&
+ \frac{1}{\sqrt{2}} \left (d^{abc} + i f^{abc} \right )
\left (\phi^{b*} \phi^c - \overline\phi^b \overline\phi^{c*} \right )
\Bigr ]
.
\eeq
The supersymmetric $F$-term contribution is
\beq
V_F &=& |F_0|^2 + |F^a|^2 + |\overline F_0|^2 + |\overline F^a|^2 ,
\eeq
where
\beq
F_0^* &=& 
\muPhi (\sqrt{3}\> \overline v + \overline\phi_0)
- y (\sqrt{3}\>  v + \phi_0)^2/\sqrt{3}
\,+\, y \phi^a \phi^a/2\sqrt{3} 
,
\\
\overline F_0^* &=& 
\muPhi (\sqrt{3}\>  v + \phi_0)
- \overline y (\sqrt{3}\>  \overline v + \overline \phi_0)^2/\sqrt{3}
\,+\, \overline y \overline\phi^a \overline\phi^a/2\sqrt{3} 
,
\\
F^{a*} &=&
\muPhi \overline\phi^a + y (v + \phi_0/\sqrt{3}) \phi^a 
-y d^{abc} \phi^b \phi^c/\sqrt{2}
,
\\
\overline F^{a*} &=&
\muPhi \phi^a +
\overline y (\overline v + \overline\phi_0/\sqrt{3}) \overline\phi^a 
-\overline y d^{abc} \overline\phi^b \overline\phi^c/\sqrt{2}
.
\eeq
Finally, the expansion of the soft supersymmetry-breaking part $V_{\rm soft}$ is
\beq
V_{\rm soft} &=&
\biggl \{
a \left [
(v + \phi_0/\sqrt{3})^3 
- \frac{1}{2} (v + \phi_0/\sqrt{3}) \phi^a \phi^a
+ \frac{1}{3\sqrt{2}} d^{abc} \phi^a \phi^b \phi^c \right ]
\nonumber \\ &&
+ \overline a \left [
(\overline v + \overline \phi_0/\sqrt{3})^3 
- \frac{1}{2} (\overline v + \overline \phi_0/\sqrt{3}) \overline \phi^a \overline \phi^a
+ \frac{1}{3\sqrt{2}} d^{abc} \overline \phi^a \overline \phi^b \overline \phi^c \right ]
\nonumber \\ &&
- \bphi \left [(\sqrt{3}\, v + \phi_0) (\sqrt{3}\, \overline v + \overline \phi_0)
+ \phi^a \overline \phi^a \right ]
\biggr \} + {\rm c.c.}
\nonumber \\ &&
+ m^2 (|\sqrt{3}\, v + \phi_0|^2 + |\phi^a|^2)
+ \overline m^2 (|\sqrt{3}\, \overline v + \overline\phi_0|^2 + |\overline\phi^a|^2).
\label{eq:Vsoftexpanded}
\eeq
 
Isolating the quadratic parts of $V$, the squared masses for the real scalar fields in $\Phi$, $\overline \Phi$ are as follows. Writing $\phi^a = (R^a + i I^a)/\sqrt{2}$ and 
$\overline \phi^a = (\overline R^a + i \overline I^a)/\sqrt{2}$
and $\phi_0 = (R_0 + i I_0)/\sqrt{2}$ and 
$\overline \phi_0 = (\overline R_0 + i \overline I_0)/\sqrt{2}$,
the singlet spin-0 squared mass matrix in the basis 
$\varphi = (R_0, \overline R_0, I_0, \overline I_0)$ is:
\beq
M_{\varphi}^2 &=&
\begin{pmatrix}
U + 4 |yv|^2 -2 X_1 & -2X_2 - {\rm Re}[b_\phi] & 2Y_1 & -2Y_2 + {\rm Im[b_\phi]} 
\\
-2X_2 - {\rm Re}[b_\phi] & 
\phantom{i}
\overline U + 4 |\overline y \overline v|^2 - 2 \overline X_1 
\phantom{i}
& 2Y_2 + {\rm Im}[b_\phi] & 2 \overline Y_1 
\\
2 Y_1 & 2Y_2 + {\rm Im}[b_\phi] & 
\phantom{i}U + 4|yv|^2 +2 X_1 
\phantom{i}
& -2 X_2 + {\rm Re}[b_\phi]
\\
-2 Y_2 + {\rm Im}[b_\phi] &  2\overline Y_1 & -2X_2 + {\rm Re}[b_\phi] & 
\phantom{i}\overline U + 4 |\overline y \overline v|^2 + 2\overline X_1
\end{pmatrix}
,
\phantom{...xx...}
\eeq
where
\beq
U &=& |\mu|^2 + m^2,
\qquad\qquad
\overline U \>=\> |\mu|^2 + \overline m^2,
\\
X_1 + i Y_1 &=& y (\mu \overline v - y v^2)^* - a v
,
\\
\overline X_1 + i \overline Y_1 &=& 
\overline y (\mu v - \overline y \overline v^2)^* - \overline a \overline v
,
\\
X_2 &=& {\rm Re}[\mu^* (y v + \overline y \overline v)]
,
\qquad\quad
Y_2 \>=\> {\rm Im}[\mu^* (y v - \overline y \overline v)]
,
\eeq
and the octet spin-0 (sgluon) squared mass matrix in the basis $S^a = 
(R^a, \overline R^a, I^a, \overline I^a)$ is:
\beq
M_{S}^2 &=&
\begin{pmatrix}
U + |yv|^2 + X_1 & X_2 - {\rm Re}[b_\phi] & -Y_1 & Y_2 + {\rm Im[b_\phi]} 
\\
X_2 - {\rm Re}[b_\phi] & 
\phantom{i}
\overline U + |\overline y \overline v|^2 + \overline X_1 
\phantom{i}
& -Y_2 + {\rm Im}[b_\phi] & -\overline Y_1 
\\
-Y_1 & -Y_2 + {\rm Im}[b_\phi] & 
\phantom{i}U + |yv|^2 - X_1 
\phantom{i}
& X_2 + {\rm Re}[b_\phi]
\\
Y_2 + {\rm Im}[b_\phi] & -\overline Y_1 & X_2 + {\rm Re}[b_\phi] & 
\phantom{i}\overline U + |\overline y \overline v|^2 - \overline X_1
\phantom{.}
\end{pmatrix}
\nonumber \\ &&
+ (g_A^2 + g_B^2) 
\begin{pmatrix}
\phantom{i}v_R^2 
& 
\phantom{i}-v_R \overline v_R\phantom{i}
& 
\phantom{i}v_R v_I\phantom{i}
& 
\phantom{i}-v_R \overline v_I
\\
-v_R \overline v_R
& 
\phantom{i}\overline v_R^2
& 
\phantom{i}-\overline v_R v_I\phantom{i}
& 
\phantom{i}\overline v_R \overline v_I
\\
v_R v_I
& 
-\overline v_R v_I 
& 
v_I^2
& 
-v_I \overline v_I
\\
-v_R \overline v_I
& 
\overline v_R \overline v_I
&
-v_I \overline v_I
&
\phantom{i}\overline v_I^2
\end{pmatrix}
,
\eeq
where
\beq
v_R + i v_I = v,
\qquad
\overline v_R + i \overline v_I = \overline v.
\eeq
The real symmetric squared mass matrices $M_{S}^2$ and $M_{\varphi}^2$ 
can be diagonalized by orthogonal transformations 
to obtained the squared mass eigenvalues for the real color octet and singlet spin-0 particles.
One of the octet spin-0 eigenvectors is the would-be Goldstone boson of the symmetry breaking,
which is absorbed as the longitudinal mode of the massive vector.
In the limit of vanishing $v_I$, $\overline v_I$, Im$[b_\phi]$, $Y_1$, and $Y_2$ 
(i.e., no CP-violating phases), the diagonalizations
separate into $2\times 2$ blocks corresponding to scalar and pseudo-scalar states.
In that case, there are two scalar and one pseudo-scalar sgluons, and two scalar and two pseudo-scalar
singlets.

\section{Minimization of the scalar potential\label{sec:minimization}}
\setcounter{equation}{0}
\setcounter{figure}{0}
\setcounter{table}{0}
\setcounter{footnote}{1}

\subsection{The supersymmetric limit\label{subsec:SUSYlimit}}

As a warm-up example and a useful limiting case, 
consider the supersymmetric limit in which the soft parameters 
$a$, $\overline a$, $b_\phi$, $m^2$, and $\overline m^2$ are all set to $0$. Then the scalar potential
as a function of $v$ and $\overline v$ becomes simply:
\beq
V(v,\overline v) = 3 |y v^2 - \mu_\Phi\overline v|^2 + 3 |\overline y\hspace{0.6pt} \overline v^2 - \mu_\Phi v|^2 ,
\eeq
as this is a $D$-flat direction. 
This has distinct minima at $v = \overline v = 0$, where the gauge symmetry is unbroken, and at
\beq
v &=& \mu_\Phi/(y^2 \overline y)^{1/3},
\quad\qquad
\overline v \>=\> \mu_\Phi/(y \overline y^2)^{1/3},
\label{eq:SUSYminVEVs}
\eeq
where the gauge symmetry is broken to $SU(3)_C$.
These are degenerate global minima, with $V=0$ in both cases, so that supersymmetry 
is not spontaneously broken. They can be checked to be minima of the 
full potential eqs.~(\ref{eq:Vtotal})-(\ref{eq:Vsoftexpanded}),
by evaluating the scalar squared masses and noting that they are non-negative, 
other than the octet of vanishing eigenvalues
corresponding to the would-be Goldstone bosons of the spontaneously broken gauge symmetry in the
case of eq.~(\ref{eq:SUSYminVEVs}). The mass spectrum of the theory contains 
a massive vector supermultiplet, consisting of the vector bosons, 
a Dirac fermion (two 2-component fermions), 
and a real scalar, all of which are octets of $SU(3)_C$ with squared masses 
$M_X^2 = (g_A^2 + g_B^2) (|v|^2 + |\overline v|^2)$. There is also a massive color octet chiral 
supermultiplet (one 2-component fermion and a complex scalar) with squared masses
\beq
M_{\rm octet}^2 = R^2 |\mu_\Phi|^2
,
\eeq
and two singlet chiral supermultiplets with squared masses
\beq
M_{\rm singlets}^2 = \left ( 2 R^2 - 3 \pm 2 R \sqrt{R^2 - 3} \right ) |\mu_\Phi|^2 ,
\label{eq:M2singletsSUSY}
\eeq
where
\beq
R = 
\left | {y}/{\overline y} \right |^{1/3} + 
\left | {\overline y}/{y} \right |^{1/3}.
\label{eq:defineRSUSY}
\eeq
Because $R \geq 2$ (with equality if $|\overline y| = |y|$), these squared masses are always
positive.
One of the singlet chiral supermultiplets is always lighter, and one always heavier, than
the octet chiral supermultiplet.
There is also an $SU(3)_C$ octet massless vector supermultiplet (the MSSM gluon and gluino),
and one massless $SU(3)_C$ octet real scalar would-be Goldstone boson which is absorbed by the
massive color octet vector boson, becoming its longitudinal mode. 

\subsection{Realistic examples with supersymmetry breaking}

Now consider the realistic case that supersymmetry breaking is included. It is useful 
to take a more general form for the possible scalar field expectation values, to include the possibility
that the remnant gauge symmetry is not $SU(3)_C$:
\beq
\langle \phi_j^k \rangle = \delta_j^k v + \delta_{j3} \delta^{k3} s,
\qquad\qquad
\langle \overline \phi_j^k \rangle = \delta_j^k \overline v + \delta_{j3} \delta^{k3} \overline s.
\eeq
Now if $s=\overline s = 0$ and $v, \overline v$ are non-zero, 
the unbroken gauge symmetry will be $SU(3)_C$. If $v = \overline v = 0$ and $s, \overline s$ are non-zero,
then the unbroken gauge symmetry is $SU(2) \times SU(2) \times U(1)$. For general $v,\overline v, s, \overline s$, the unbroken gauge symmetry would be $SU(2) \times U(1)$. I do not consider 
even more general VEVs, for which the
unbroken gauge symmetry would be even smaller.  
This is because both the $D$-term and 
$F$-term contributions to the potential are non-negative, and they favor the larger unbroken symmetries
$SU(3)_C$ or $SU(2) \times SU(2) \times U(1)$.
As found \cite{Bai:2017zhj} in the non-supersymmetric case with
one $({\bf 3}, {\bf \overline 3})$ scalar field, no local minimum is expected 
in the case of a $SU(2) \times U(1)$ or smaller residual symmetry, and I have confirmed this in numerical examples, although I have not attempted a formal or general proof.

The scalar potential $D$-term, $F$-term, and soft contributions are then:
\beq
V_D &=& 
\frac{1}{6} (g_A^2 + g_B^2) 
\left (|v + s|^2 - |\overline v + \overline s|^2 - |v|^2 + |\overline v|\right )^2
,
\\
V_F &=& 
|y v^2 - \mu_\Phi (\overline v + \overline s)|^2 
+ |\overline y \hspace{0.5pt} \overline v^2  - \mu_\Phi (v + s)|^2 
+ 2 |y v (v+s) - \mu_\Phi \overline v|^2
\nonumber \\
&&
+ 2 |\overline y\hspace{0.5pt} \overline v (\overline v+ \overline s)  - \mu_\Phi v|^2
,
\\
V_{\rm soft} &=& \left ( a v^2 (v+s) + \overline a \overline v^2 (\overline v + \overline s)
- b_\phi \left [ (v + s)(\overline v + \overline s) + 2 v \overline v \right ] \right )
+ {\rm c.c.}
\nonumber \\
&& + m^2 \left (|v+s|^2 + 2 |v|^2 \right ) 
+ \overline m^2 \left (|\overline v + \overline s|^2 + 2 |\overline v|^2 \right ) 
.
\eeq
There is a $D$-flat direction $|\overline s| = |s|$ when $\overline v = v = 0$.
This is unaffected by the $y, \overline y, a, \overline a$ couplings, but it is lifted by the $F$-term contribution $V = |\mu_\Phi|^2 (|s|^2 + |\overline s|^2)$, 
so it is not a minimum of the supersymmetric limit of the previous section. However, it can be favored
by the soft supersymmetry breaking squared mass terms, 
leading to a runaway unbounded from below (UFB) direction, 
in which $|\overline s| = |s|$  becomes arbitrarily large 
and the phase of $s \overline s$ is the same as that of $b_\phi^*$. This will occur unless
\beq
|b_\phi|
&<&   
|\mu_\Phi|^2 + (m^2 + \overline m^2)/2 
.
\label{eq:noUFBconstraint}
\eeq
This UFB solution can be separated by a barrier from other local minima with non-zero 
$v$, $\overline v$, which could therefore in principle be viable if the tunneling rate is small enough.

Next, take the possibility that $v = \overline v = 0$ with $|s|\not=|\overline s|$, 
which would lead to the symmetry breaking pattern
$SU(3)_A \times SU(3)_B \rightarrow SU(2) \times SU(2) \times U(1)$, and consider 
\beq
V(s,\overline s) &=& \frac{1}{6} (g^2_A + g^2_B) (|s|^2 - |\overline s|^2)^2 
+ (m^2 + |\mu_\Phi|^2) |s|^2
+ (\overline m^2 + |\mu_\Phi|^2) |\overline s|^2
- (b_\phi s \overline s + {\rm c.c.})
.
\label{eq:Vssbar}
\phantom{xxx}
\eeq 
Assuming $m^2\leq \overline m^2$ without loss of generality [otherwise the discussion goes through 
with $(s, m^2) \leftrightarrow (\overline s, \overline m^2)$],
the possible nontrivial stable minimum of 
$V(s, \overline s)$ is at:
\beq
|s|^2 &=& \frac{3}{4(g_A^2 + g_B^2) d} \left [
(|\mu_\Phi|^2 + \overline m^2)^2 - (|\mu_\Phi|^2 + m^2 + d)^2
\right ] 
,
\label{eq:s221}
\\
|\overline s|^2 &=& \frac{3}{4(g_A^2 + g_B^2) d} \left [
(|\mu_\Phi|^2 + \overline m^2 -d)^2 - (|\mu_\Phi|^2 + m^2)^2
\right ] ,
\label{eq:sbar221}
\eeq
where
\beq
d = \sqrt{(2 |\mu_\Phi|^2 + m^2 + \overline m^2)^2 - 4 |b_\phi|^2}.
\label{eq:d221}
\eeq
To satisfy the necessary conditions that $d$ and $|s|^2$ and $|\overline s|^2$ are real and positive, 
$|b_\phi|$ must satisfy:
\beq
(|\mu_\Phi|^2 + m^2)(|\mu_\Phi|^2 + \overline m^2) \><\> 
|b_\phi|^2
\><\>
(|\mu_\Phi|^2 + m^2)(|\mu_\Phi|^2 + \overline m^2) + \frac{1}{4} (m^2 - \overline m^2)^2,
\label{eq:brange221}
\eeq
where the right inequality coincides with the no-UFB condition eq.~(\ref{eq:noUFBconstraint}),
and the left inequality coincides with the destabilization of the trivial vacuum with
$s=\overline s = 0$.
In practice, this is usually a very narrow range of allowed $|b_\phi|$; in particular, it
vanishes in the limit $m^2 = \overline m^2$. Also,
while the condition eq.~(\ref{eq:brange221}) is necessary and sufficient for a non-trivial minimum
of $V(s,\overline s)$, it is far from sufficient to guarantee that 
eqs.~(\ref{eq:s221})-(\ref{eq:d221}) provide
a local minimum of the whole potential
(not restricted to the $s,\overline s$ subspace).
The sufficient conditions follow from also requiring the positivity of the $36-9=27$ non-Goldstone
squared mass eigenvalues, of which 8 are distinct. 
These depend on the other parameters in a more complicated way, and can be evaluated on a case-by-case
basis. 

Now consider the 
$D$-flat direction defined by $\overline s = s = 0$ and non-zero $v,\overline v$, which gives 
$SU(3)_A \times SU(3)_B \rightarrow SU(3)_C$ as desired. For simplicity, 
consider first a special case that
has $\Phi \leftrightarrow \overline \Phi$ and CP symmetries, where
$\overline y = y$ is real and positive by convention, $\mu_\Phi$ is chosen to be real and positive,
$b$ and $\overline a = a$ are chosen to be real but not necessarily positive, 
and\footnote{This choice precludes the possibility of 
a $SU(2) \times SU(2) \times U(1)$-preserving minimum, as just discussed.}
 $\overline m^2 = m^2$, which must be real (by the reality of the Lagrangian)
but not necessarily positive.
For convenience, define real
dimensionless supersymmetry breaking parameters
\beq
A &=& a/(y \mu_\Phi),
\label{eq:defineA}
\\
B &=& b/\mu_\Phi^2,
\label{eq:defineB}
\\
C &=& m^2/\mu_\Phi^2,
\label{eq:defineC}
\eeq 
in terms of which 
eq.~(\ref{eq:noUFBconstraint}) becomes the requirement 
\beq
|B| < 1 + C
\label{eq:nospecialUFB}
\eeq 
to avoid an UFB runaway solution. 
Then one can look for minima 
\beq
v = (\mu_\Phi/y) x e^{i \alpha}, \qquad
\overline v = (\mu_\Phi/y) x e^{i \beta},
\label{eq:vxalpha}
\eeq 
where $x$ is real, non-negative, and dimensionless, and $\alpha$ and $\beta$ are phases.
By examining the first derivatives of the potential, one finds that a minimum with
$x\not=0$ that satisfies eq.~(\ref{eq:nospecialUFB}) must have $\beta = \alpha$.
The potential then becomes simply
\beq
V(x,\alpha) &=& \frac{6 \mu_\Phi^4}{y^2}\> x^2\! \left (
x^2 +  \left [\left(\frac{8}{3} \cos^2\alpha -2 \right ) A - 2 \right ] x \cos\alpha
+ B \left (1 - 2 \cos^2\alpha \right ) + C + 1 
\right ).\phantom{xxx}
\eeq 
Minimizing the restricted potential $V(x,\alpha)$ gives a necessary condition, but one
must also check using the full scalar potential 
that at any putative local minimum, 
all of the 36 real scalar squared masses are non-negative, including an octet of vanishing  
scalar squared masses for the would-be Goldstone bosons.

For $A=B=C=0$, one recovers the supersymmetric limit of the previous section, with a minimum at $x=1$,
$\alpha=0$. More generally, the supersymmetric part of the scalar potential clearly favors $\alpha=0$
when $x\not=0$. 
The $A$ term also favors $\alpha=0$ for large negative $A$. However, for large positive $A$, the $A$
term part favors symmetry breaking with $\cos^2\alpha = 1/2$. The $B$ term favors 
$\alpha=0$ if $B<0$, but $\alpha = \pi$ if $B>0$.
The tension between these contributions means that even though all potential parameters were
chosen to be real, the VEV can be forced to be complex at a local minimum
if $A$ is positive and sufficiently large. There are thus two types of possible local
symmetry breaking minima, which from now on are parameterized by $x e^{i\alpha} = x_R + i x_I$.
Without loss of generality, one can take $x_I$ to be non-negative. 

For the first type, the VEV is real and
satisfies the stationary condition
\beq
2 x_R^2 + (A-3) x_R + 1 + C- B &=& 0,
\label{eq:realVEVextremum}
\eeq
leading to
\beq
x_R &=& \frac{3-A}{4} \left (1 + \sqrt{1 + 8 (B-C-1)/(3-A)^2} \right ) ,
\label{eq:realVEVxR}
\\
x_I &=& 0,
\eeq 
For this to be a local minimum, it is necessary but not sufficient that the argument of the
square root is positive:
\beq 
(3 -A)^2 &>& 8 (1 -B +C). 
\label{eq:type1req}
\eeq
From requiring positivity of the $36-8 = 28$ non-Goldstone scalar squared masses, 
one finds the other necessary conditions:
\beq
(1 + n x_R)^2 + C &>& |B + n x_R (A + x_R -1)|
\label{eq:type1req3}
\eeq
to be imposed for each of $n=1,2,-2$,
and
\beq
(g_A^2 + g_B^2) x_R^2 + (1 - x_R)^2 &>& -C.
\label{eq:type1reqgauge}
\eeq 
Together, the five conditions~(\ref{eq:type1req})-(\ref{eq:type1reqgauge})
are sufficient to guarantee the existence of this local minimum.
The constraint (\ref{eq:type1reqgauge}) is the only one that depends on the gauge 
couplings, and it rarely comes into play; it is automatic unless $C<0$, 
and even then it is always satisfied for sufficiently large gauge couplings.
If the no-UFB condition eq.~(\ref{eq:nospecialUFB}) is also imposed, then the three
conditions of eq.~(\ref{eq:type1req3})
can be simplified to:
\beq
2 x_R + B &>& 0,
\\
(1 - 3 A) x_R + 2 B &>& 0,
\\
(11 - A) x_R -2 + 4B - 2C &>& 0.
\eeq
For eq.~(\ref{eq:realVEVxR}) to be the global minimum, 
it is necessary but not sufficient (because of the possibility of the
second type of solution below) that eq.~(\ref{eq:nospecialUFB}) is 
also satisfied as well as $V\leq 0$, which yields 
\beq
(3-A)^2 &\geq& 9 (1-B+C),
\label{eq:type1reqglobal}
\eeq
which is slightly stronger than eq.~(\ref{eq:type1req}).

The second type of local minimum has a complex VEV, with stationary conditions
\beq
x_I^2 + x_R^2 - (1+A) x_R + (1+B+C)/2 &=& 0,
\\
4 A x_R^2 - [(1+A)^2 + 2 B] x_R + (1+A)(1+B+C)/2 &=& 0,
\eeq
leading to
\beq
x_R &=& 
\frac{(1+A)^2 + 2 B}{8A} 
\left [1 + \sqrt{1 -8 A (1+A)(1 + B + C)/[(1 + A)^2 + 2 B]^2 }\right ],
\label{eq:xRcomplex}
\\ 
x_I &=& \sqrt{x_R (1 + A - x_R) - (1 + B + C)/2}
.
\label{eq:xIcomplex}
\eeq
As necessary but not sufficient requirements,
both square roots must have positive argument, so 
\beq
[(1 + A)^2 + 2 B]^2 &>& 8 A (1+A)(1 + B + C),
\label{eq:type2req1}
\\
2 x_R (1 + A - x_R) &>&  1 + B + C,
\eeq
The remaining necessary conditions, coming from positivity of the four distinct non-Goldstone 
scalar boson squared mass eigenvalues, are:
\beq
(1+ n x_R)^2 + n^2 x_I^2 + C &>&
        \sqrt{[B - n x_I^2 + n x_R (A + x_R - 1)]^2 + n^2 x_I^2 (1 + A - 2 x_R)^2}
,
\phantom{xxx}
\label{eq:type2req3}
\eeq
again for each of $n=1,2,-2$,
and
\beq
(g_A^2 + g_B^2) (x_I^2 + x_R^2) + (1 - x_R)^2 + x_I^2 &>& -C.
\label{eq:type2reqgauge}
\eeq
Together, the six conditions 
eqs.~(\ref{eq:type2req1})-(\ref{eq:type2reqgauge}) 
are sufficient to guarantee the existence of this local minimum.
Again, eq.~(\ref{eq:type2reqgauge}) can only come into play if $C<0$, and even then 
it is automatically satisfied if the gauge couplings are sufficiently large.
For a local minimum of this type
to be the global minimum, 
it is necessary but not sufficient (because of the possibility of the
first type of solution described above) that eq.~(\ref{eq:nospecialUFB}) is 
satisfied as well as $V\leq 0$, a constraint that can be written as
\beq
(x_R^2 + x_I^2)^2 + (2 B - 8 A x_R/3) x_R^2 &\geq& 0.
\label{eq:globalminconstraintcomplex}
\eeq

\begin{figure}[t]
\begin{center}
\begin{minipage}[]{0.49\linewidth}
\includegraphics[width=\linewidth,angle=0]{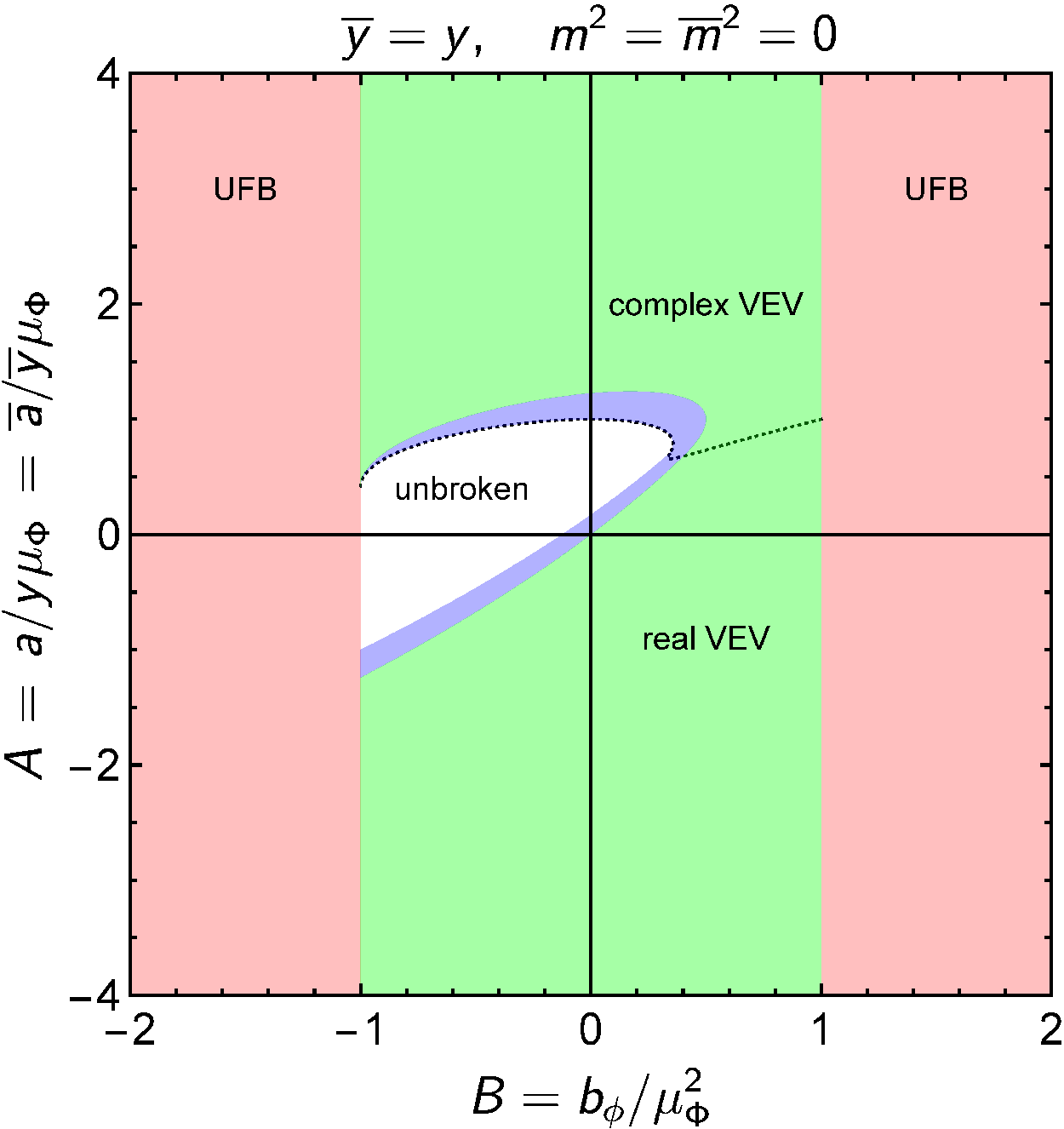}
\end{minipage}
\begin{minipage}[]{0.49\linewidth}
\includegraphics[width=\linewidth,angle=0]{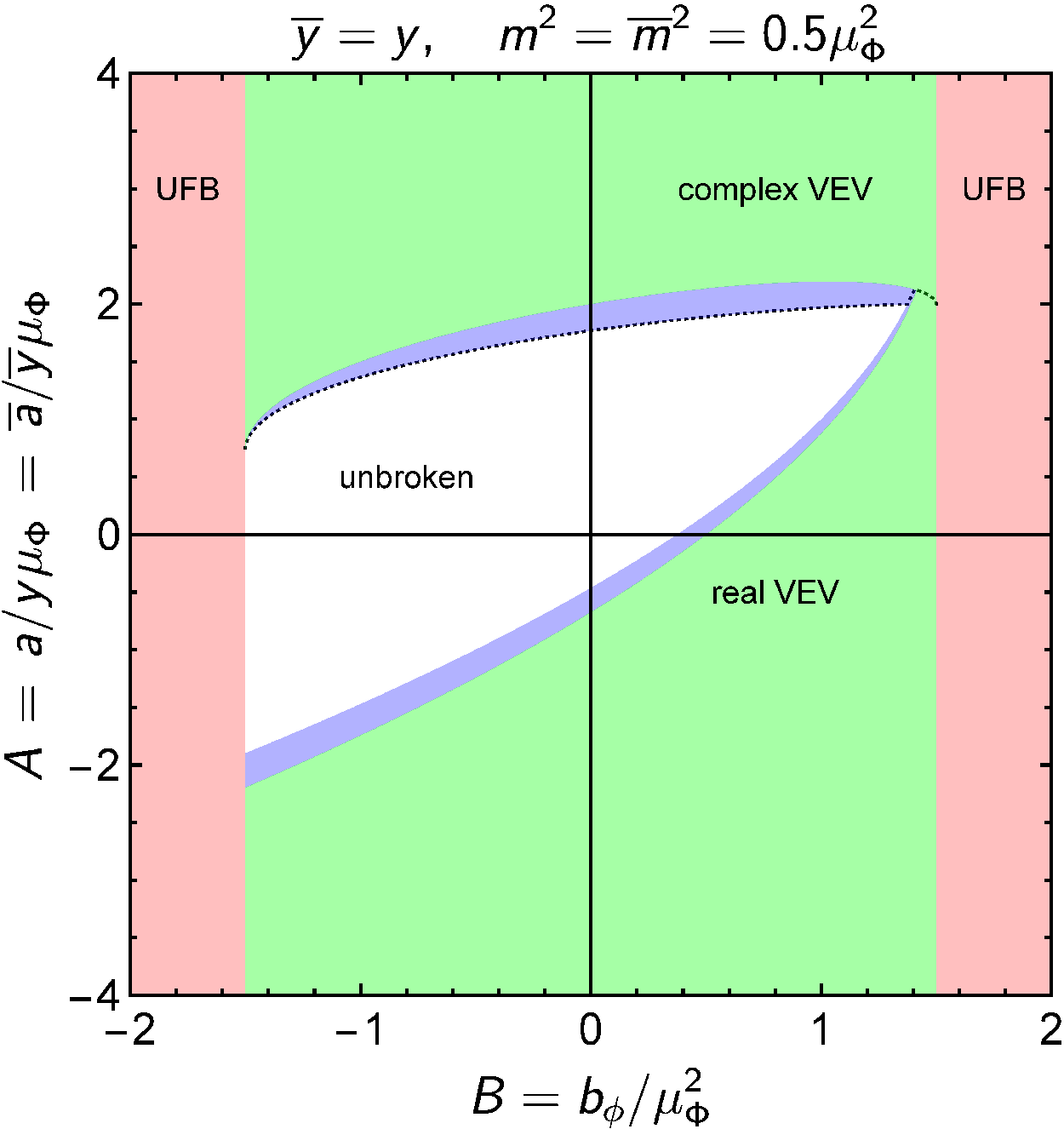}
\end{minipage}
\end{center}

\vspace{-0.5cm}

\caption{\label{fig:phase}
Phase diagrams for symmetry breaking in the case $\overline y = y$
and $\overline a = a$, for $m^2 = \overline m^2 = 0$ (left panel)
and $m^2 = \overline m^2 = 0.5 \mu^2_{\Phi}$ (right panel), with the input parameters 
$\mu_\Phi$ and $y$ real and positive, and  $a, b_\phi$ chosen real, so that the discussion of
eqs.~(\ref{eq:nospecialUFB})-(\ref{eq:globalminconstraintcomplex}) 
applies to the minimization of the scalar potential.
In the red shaded regions on the left and right sides of each 
plot, the scalar potential has an
unbounded from below direction. 
In the unshaded central region, the $SU(3)_A \times SU(3)_B$ gauge 
symmetry is not broken at any local minimum of the potential. The symmetry breaking
$SU(3)_A \times SU(3)_B \rightarrow SU(3)_C$ occurs at a global minimum 
of the potential in the large green shaded regions, and at only a 
local minimum in the thin blue shaded regions.
The supersymmetric limit occurs at the origin $(b_\phi, a) = (0,0)$ 
in the left panel; there
the local symmetry breaking minimum is degenerate with 
the local non-symmetry breaking minimum.
In each panel, the lowest $SU(3)_C$-symmetric minimum has complex 
$v=\overline v$ above the dotted curve.
}
\end{figure}

The implications of the preceding results are illustrated in Figure \ref{fig:phase}, 
which shows a phase diagram for symmetry breaking in the $B=b_\phi/\mu_\Phi^2$ vs.~$A = a/y\mu_\Phi = 
\overline a/\overline y\mu_\Phi$ 
plane, for the choices $C= m^2/\mu_\Phi^2 = \overline m^2/\mu_\Phi^2 = 0$ 
(left panel) and $0.5$ (right panel).
As noted above, there can be no $SU(2)\times SU(2) \times U(1)$-preserving vacuum here, because of the
choice $m^2 = \overline m^2$. 
The red shaded regions on the left and right sides of each plot have UFB runaway solutions because
$|B|$ is too large. In the central unshaded regions,
there are no symmetry breaking local minima.
The green region shows the points where the global
minimum of the potential
breaks $SU(3)_A \times SU(3)_B \rightarrow SU(3)_C$, and the blue region shows where there is
at least one such local minimum with no UFB runaway. These are the regions that could be our world.
At $A=B=C=0$, the supersymmetric limit is realized, so that this point is on the border
between the local and global minimum regions in the left panel. A dotted curve separates the region
where the lowest symmetry breaking local minimum has a real VEV 
from the region where it has a complex VEV
(which occurs for $A$ positive and not too small), 
given our choice of all real input parameters.

\begin{figure}[t]
\begin{center}
\begin{minipage}[]{0.495\linewidth}
\includegraphics[width=\linewidth,angle=0]{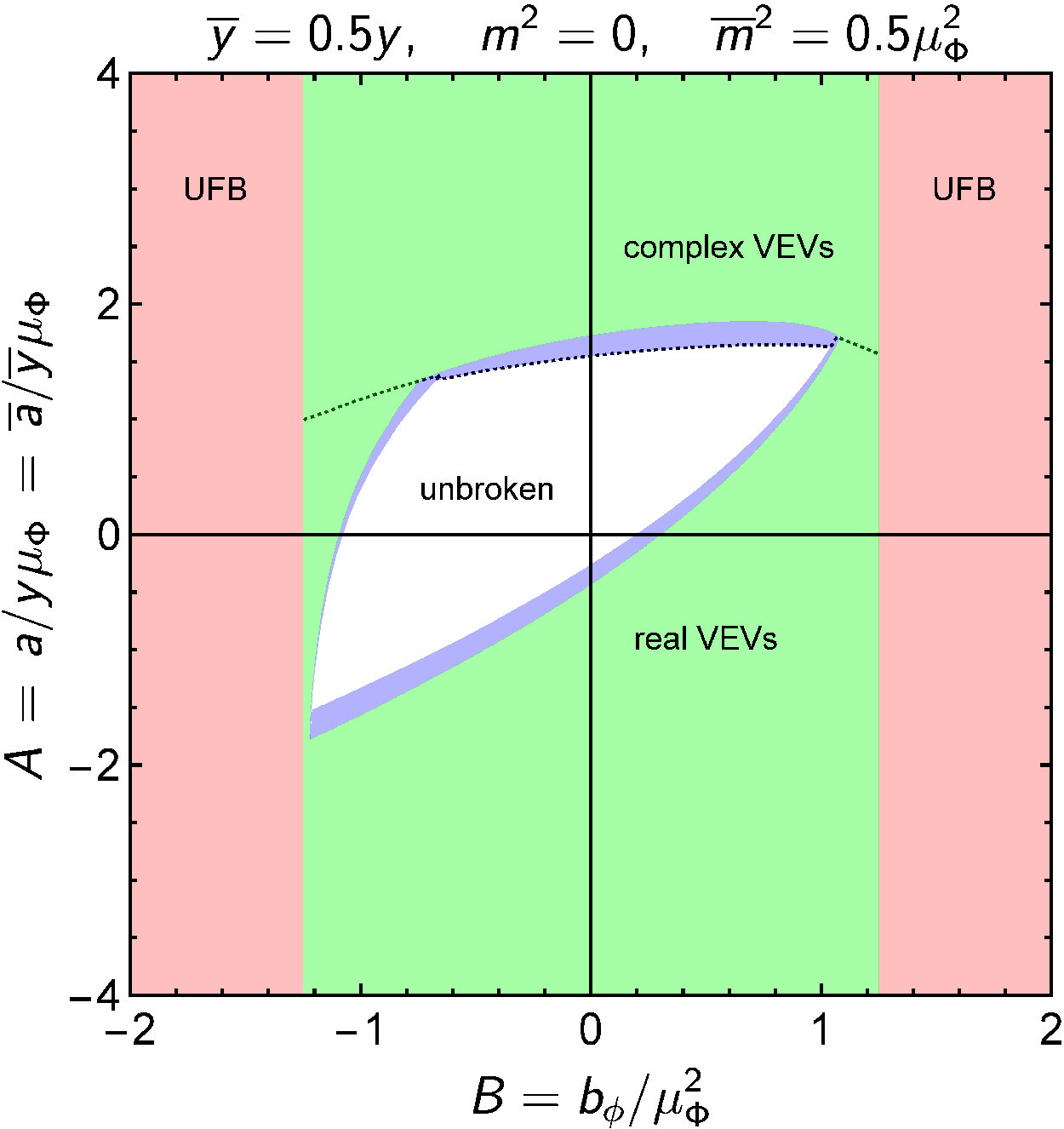}
\end{minipage}
\begin{minipage}[]{0.495\linewidth}
\includegraphics[width=\linewidth,angle=0]{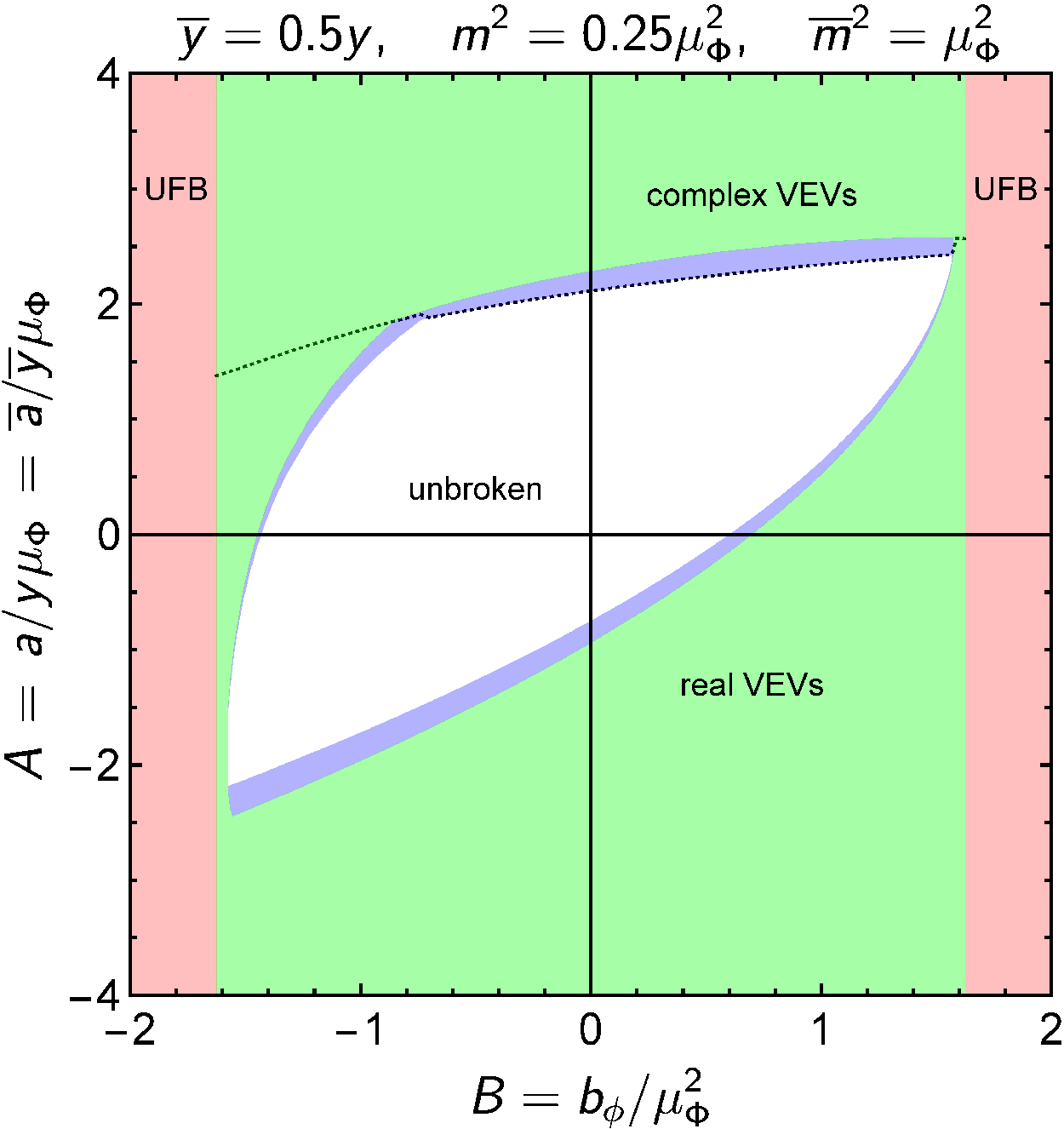}
\end{minipage}
\end{center}

\vspace{-0.5cm}

\caption{\label{fig:phasetwo}
Phase diagrams for symmetry breaking in the case $\overline y = 0.5 y$ real and positive,
with $m^2 = 0$, $\overline m^2 = 0.5 \mu_\Phi^2$ (left panel) and with
$m^2 = 0.25 \mu^2_{\Phi}$, $\overline m^2 = \mu_\Phi^2$ (right panel).
The axes are $B=b_\phi/\mu_\Phi^2$ and $A = a/y\mu_\Phi = \overline a/\overline y\mu_\Phi$, 
which are assumed to be real but can have either sign. 
In the red shaded regions on the left and right sides of each 
plot, the scalar potential has an unbounded from below direction. 
In the unshaded central region, the $SU(3)_A \times SU(3)_B$ gauge 
symmetry is not broken at any local minimum of the potential. The symmetry breaking
$SU(3)_A \times SU(3)_B \rightarrow SU(3)_C$ occurs at a global minimum 
of the potential in the large green shaded regions, and at only a 
local minimum in the thinner blue shaded regions.
In each panel, the lowest $SU(3)_C$-symmetric minimum has complex VEVs 
$v$ and $\overline v$ above the dotted curve.
Minima with unbroken gauge group $SU(2) \times SU(2) \times U(1)$ 
can also occur, but only in very small regions that are subsets of 
thin strips adjacent to the UFB region, namely
$1.2247 < |B| < 1.25$ (left panel)
and $1.5811 < |B| < 1.625$ (right panel). These small regions are not shown 
because they 
depend on the  other parameters.}
\end{figure}

In view of the rather complicated set of requirements 
given above even in the simplifying 
case of assumed real parameters with a $\Phi \leftrightarrow \overline \Phi$ 
symmetry in the Lagrangian, I have not attempted to characterize the necessary 
and sufficient conditions in the general case.
However, using numerical methods 
I have checked that in generic cases, for large areas in a general parameter space, there are 
global minima that realize the $SU(3)_A \times SU(3)_B \rightarrow SU(3)_C$ breaking.
For example, Figure \ref{fig:phasetwo}
shows phase diagrams for the 
case that there is no symmetry between $\Phi$ and $\overline \Phi$, for
$\overline y = 0.5 y$ real and positive and with
$m^2 = 0$, 
$\overline m^2 = 0.5 \mu^2_{\Phi}$ (left panel) 
and with
$m^2 = 0.25 \mu^2_{\Phi}$, 
$\overline m^2 = \mu^2_{\Phi}$ (right panel).
The axes of the plots are $b_\phi/\mu_\Phi^2$ and $a/y\mu_\Phi = \overline a/\overline y\mu_\Phi$, 
which are assumed to be real but can have either sign. 
In this example, because $m^2 \not= \overline m^2$, 
there are very small regions where minima with unbroken gauge group $SU(2) \times SU(2) \times U(1)$ 
can exist, depending on the other parameters. 
From eq.~(\ref{eq:brange221}), these occur within the narrow ranges adjacent to the UFB region,
$1.2247 < |B| < 1.25$ (left panel)
and $1.5811 < |B| < 1.625$ (right panel). The exact extents of these small regions depend
on other parameters besides the plot axes, so they are not shown.
I have also checked in other examples that global
minima with unbroken gauge group $SU(3)_C$ do occur in large regions of
generic parameter space, including where $\mu_\Phi$ and the soft input
parameters are allowed to have complex phases, and that smaller residual gauge symmetries
like $SU(2) \times U(1)$ generally do not occur.


\section{Model realization with gauge coupling unification\label{sec:realmodel}}
\setcounter{equation}{0}
\setcounter{figure}{0}
\setcounter{table}{0}
\setcounter{footnote}{1}

\subsection{Renormalization group running\label{sec:RGEs}}

One aspect of low-energy supersymmetry that has often been touted as an attractive feature is
the apparent unification of gauge couplings above $10^{16}$ GeV. In the case that $SU(3)_C$ 
is the remnant of 
two independent $SU(3)$ gauge groups, this is certainly no longer automatic (but as we will see it can at least be accommodated). Furthermore, given the Standard Model value of $\alpha_S$, the formula eq.~(\ref{eq:gCfromgAgB}) implies
that both $g_{A}$ and $g_{B}$ must be fairly strong at the multi-TeV scale, since $g_{3}$ 
is necessarily smaller than both of them, and if they were equal, $g_{3} \approx g_{A}/\sqrt{2}$.
Assuming that the MSSM quark supermultiplets live in the $SU(3)_A$ representation, then the presence
of 6 additional triplets $\Phi$ and $\overline\Phi$ 
means that the 1-loop $\beta$ function for $SU(3)_A$
must be non-negative, and the 2-loop $\beta$ function is positive,
so that $g_{A}$ cannot be asymptotically free as in the MSSM. To unify with 
$g_{B}$, additional fields charged under $SU(3)_B$ must be included.
 
There are many ways to include chiral superfield representations that are charged under $SU(3)_B$. Suppose that there are additional vectorlike quark and lepton
supermultiplets in representations (and their conjugates) as in the MSSM, but with color charges only
under $SU(3)_B$. This will ensure that the new fields are not exotic and 
none of them need be stable, since they can decay by mixing 
with MSSM states. In particular, consider  
possible chiral supermultiplets in the following
vectorlike representations of
$SU(3)_A \times SU(3)_B \times SU(2)_L \times U(1)_Y$: 
\beq
n_Q &\times& \Bigl [({\bf 1}, {\bf 3}, {\bf 2}, \frac{1}{6}) + 
({\bf 1}, {\bf \overline 3}, {\bf 2}, -\frac{1}{6}) \Bigr ] 
,\\
n_d &\times& \Bigl [({\bf 1}, {\bf 3}, {\bf 1}, -\frac{1}{3}) + 
({\bf 1}, {\bf \overline 3}, {\bf 1}, \frac{1}{3}) \Bigr] 
,\\
n_u &\times& \Bigl [({\bf 1}, {\bf 3}, {\bf 1}, \frac{2}{3}) + 
({\bf 1}, {\bf \overline 3}, {\bf 1}, -\frac{2}{3}) \Bigr] 
,\\
n_L &\times& \Bigl [({\bf 1}, {\bf 1}, {\bf 2}, -\frac{1}{2}) + 
({\bf 1}, {\bf 1}, {\bf 2}, \frac{1}{2}) \Bigr] 
,\\
n_e &\times& \Bigl [({\bf 1}, {\bf 1}, {\bf 1}, -1) + 
({\bf 1}, {\bf 1}, {\bf 1}, 1) \Bigr],
\eeq
for integers
$n_Q$, $n_d$, $n_u$, $n_L$, and $n_e$.
These fields are supposed to 
have weak isosinglet bare masses in the multi-TeV range, due to whatever mechanism also provides for the MSSM $\mu$ term. They can also mix with the MSSM
quarks and leptons, in the case of quarks through Yukawa couplings 
to $\Phi$ and $\overline \Phi$. That mixing is assumed here to be too small to affect anything else
significantly.
In the following, beta functions will be denoted in the general loop expansion form
\beq
\beta_X = \sum_{n \geq 1} \frac{1}{(16\pi^2)^n} \beta_X^{(n)}.
\eeq
Then at 2-loop order,%
\footnote{In all numerical results below, 
the full 3-loop beta functions are used to run all supersymmetric parameters
and the 2-loop results are used for soft parameters. These 
can be straightforwardly obtained from
the general expressions in refs.~\cite{Jones:1974pg,Jones:1983vk, West:1984dg, Parkes:1984dh,
Martin:1993yx, Martin:1993zk, Yamada:1994id, Jack:1994kd, Jack:1994rk,
Jack:1996qq, Jack:1996vg, Jack:1996cn, Jack:1997pa, Jack:1997eh,
Jack:1998iy}, 
so only the partial 2-loop or 1-loop formulas are shown here for illustration.} 
the gauge couplings in a Grand Unified Theory (GUT) normalization have beta functions:
\beq
\beta_{g_{A}}^{(1)} &=& 0
,
\\
\beta_{g_{A}}^{(2)} &=& 
g_{A}^3 \Bigl (48 g_{A}^2 + 16 g_{B}^2 + 9 g_2^2 + \frac{11}{5} g_1^2
- 6 y^2 - 6 \overline y^2 - 4 y_t^2 - 4 y_b^2 \Bigr )
,
\\
\beta_{g_{B}}^{(1)} &=& g_{B}^3 (-6 + 2 n_Q + n_u + n_d)
,
\\
\beta_{g_{B}}^{(2)} &=& 
g_{B}^3 \Bigl ([-20 + \frac{34}{3} (2 n_Q + n_d + n_u)] g_{B}^2 + 16 g_{A}^2 
+ 6 n_Q g_2^2 
\nonumber \\ &&
+ \frac{2}{15} [n_Q + 2 n_d + 8 n_u] g_1^2 - 6 y^2 - 6 \overline y^2
\Bigr )
,
\\
\beta_{g_{2}}^{(1)} &=& g_2^3 \left (1 + 3 n_Q + n_L \right ),
\\
\beta_{g_{2}}^{(2)} &=& g_2^3 \Bigl (
24 g_{A}^2 + 16 n_Q g_{B}^2 + [25 + 21 n_Q + 7 n_L] g_2^2
+ \frac{1}{5} [9 + n_Q + 3 n_L] g_1^2
\nonumber \\ &&
-6 y_t^2 - 6 y_b^2 - 2 y_\tau^2
\Bigr ) ,
\\
\beta_{g_{1}}^{(1)} &=& \frac{g_1^3}{5} 
\left (33 + n_Q + 2 n_d + 8 n_u + 3 n_L + 6 n_3 \right ),
\\
\beta_{g_{1}}^{(2)} &=& \frac{g_1^3}{5} \Bigl (
88 g_{A}^2 + \frac{16}{3} [n_Q + 2 n_d + 8 n_u] g_B^2
+ [27 + 3 n_Q + 9 n_L] g_2^2
\nonumber \\ &&
+ \frac{g_1^2}{15} [597 + n_Q + 8 n_d + 128 n_u + 27 n_L + 216 n_e]
- 26 y_t^2 - 14 y_b^2 - 18 y_\tau^2
\Bigr ) .
\eeq
The $SU(3)_A$ coupling does not run in the 1-loop approximation, but this is an accident,
violated by 2-loop effects.
The Yukawa couplings $y$ and $\overline y$ will be assumed not to be small in the following,
and so their running is important, and given by:
\beq
\beta_y^{(1)} &=& y (6 y^2 - 8 g_{A}^2 - 8 g_{B}^2 ),
\label{eq:betayone}
\\
\beta_y^{(2)} &=& 8 y \Bigl (
\frac{8}{3} g_{A}^4 + \frac{16}{3} g_{A}^2 g_{B}^2 
+ \left [ 2 n_{Q} + n_d + n_u -\frac{10}{3} \right ] g_{B}^4
+ 2 (g_{A}^2 + g_{B}^2) y^2
-3 y^4  
\Bigr )
,
\eeq
with the same equations for  $y \rightarrow \overline y$.
The beta functions for the top-quark, bottom-quark, and tau-lepton Yukawa couplings
are obtained from the MSSM results with the replacement $g_3 \rightarrow g_{A}$.

There are several choices for the integers
$n_Q$, $n_d$, $n_u$, $n_L$, and $n_e$ that can lead to approximate gauge coupling unification.
In the following, I will simply choose one that seems interesting, with no claim 
or expectation of uniqueness:
\beq
n_Q = 1,\qquad n_d = 3,\qquad n_u = 0,\qquad n_L = 0,\qquad n_e = 1.
\eeq 
It is also possible, for example, to include a chiral supermultiplet which would transform
as an octet under $SU(3)_B$; this would also lead to three new possible Yukawa couplings.
One reason for the choice made here is that 
one can arrange for gauge coupling unification at high scales
while having $g_B > g_A$ at the symmetry breaking scale, 
with $g_3$ consistent with the Standard Model QCD coupling.

Since $\beta_{g_{A}}^{(1)} = 0$ and $\beta_{g_{B}}^{(1)} = -g_{B}^2$ are both accidentally
small in magnitude due to the choice of chiral superfield representations, 
and $\beta_{g_{A}}^{(2)}$ and $\beta_{g_{B}}^{(2)}$ both have large positive contributions,
the RG running can have a character similar to the Caswell-Banks-Zaks infrared fixed point 
\cite{Caswell:1974gg,Banks:1981nn}, 
although here the conformal regime is not actually reached. 
In the following, I consider a case that realizes approximate
gauge coupling unification through $y$ and $\overline y$ that are large at the TeV scale. 
This is natural in the sense that the negative contributions 
proportional to $g_{A}^2$ and $g_{B}^2$ 
in eq.~(\ref{eq:betayone}) will drive $y$ and $\overline y$ to be larger in
the infrared, since the $SU(3)$ gauge couplings are necessarily large. However, when
$y$ and $\overline y$ themselves 
become sufficiently large, the terms proportional to $y^2$ 
in eq.~(\ref{eq:betayone}) and proportional to $\overline y^2$ in its counterpart for $\overline y$ 
will put the brakes on, leading to a quasi-fixed point behavior. (This is not a true fixed point,
because $g_{A}$ and $g_{B}$ are still running, and thus provide a moving target.)

\begin{figure}[t]
\begin{center}
\begin{minipage}[]{0.495\linewidth}
\includegraphics[width=\linewidth,angle=0]{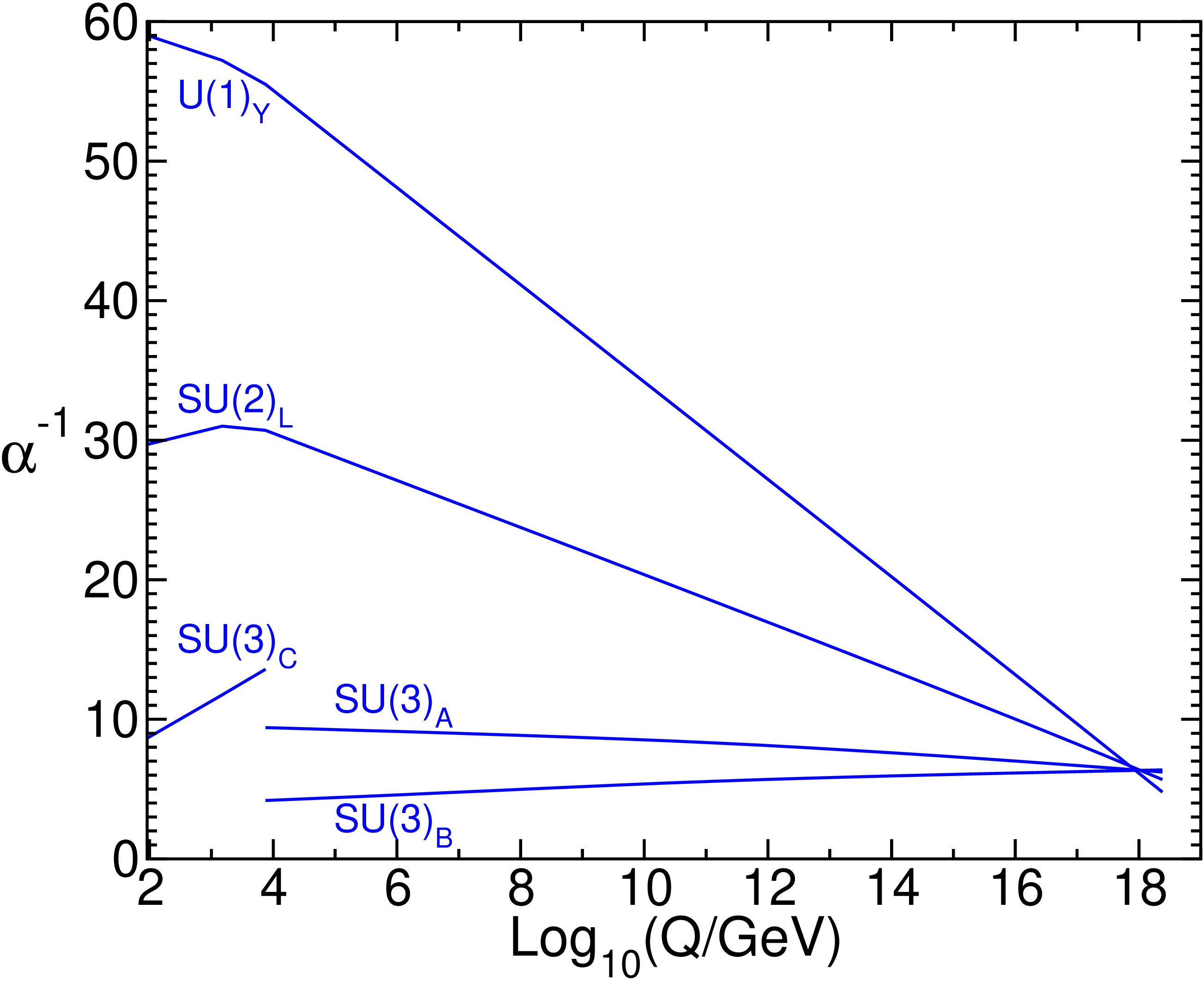}
\end{minipage}
\begin{minipage}[]{0.495\linewidth}
\includegraphics[width=\linewidth,angle=0]{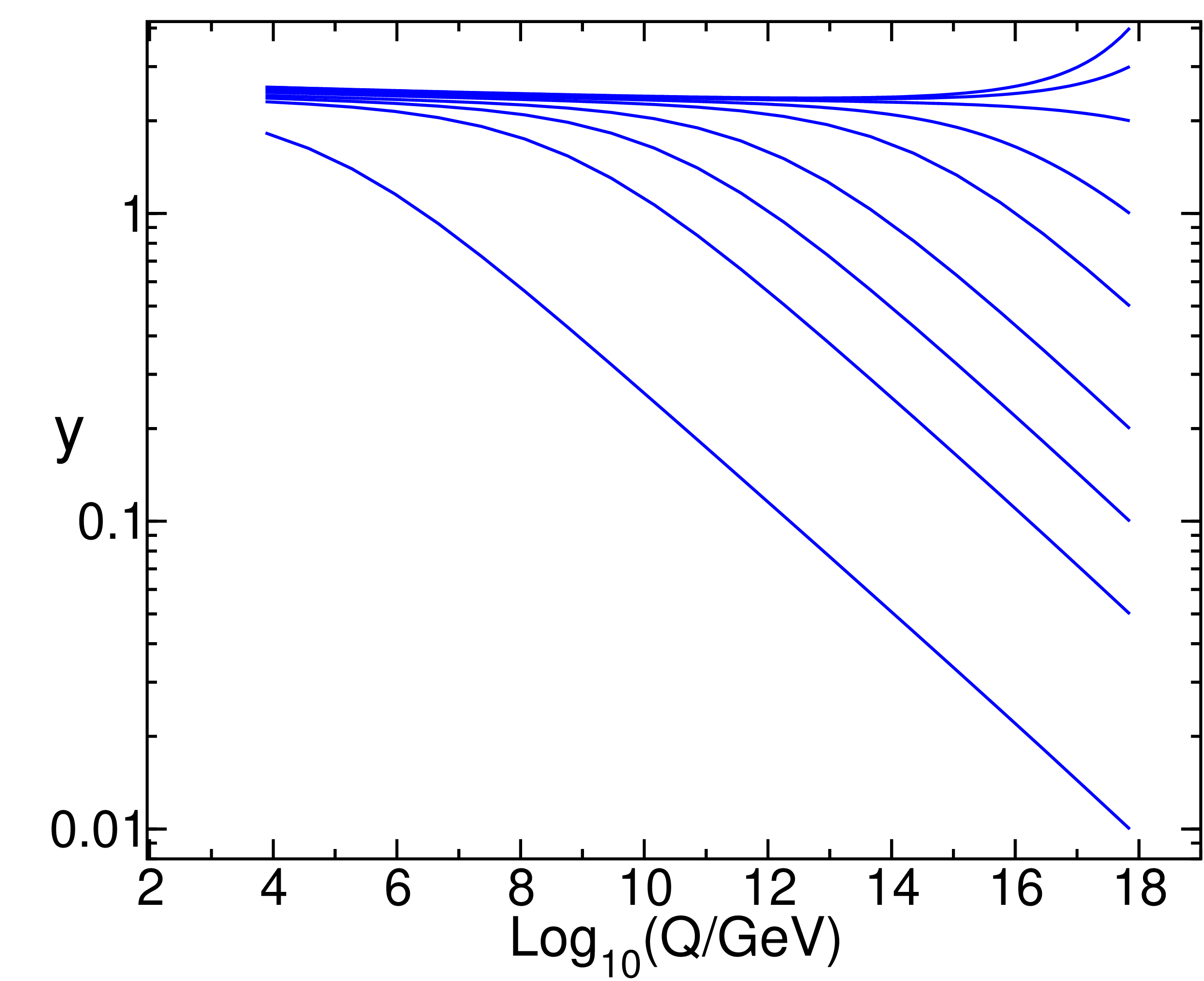}
\end{minipage}
\end{center}
\caption{\label{fig:RGrunning}
Three-loop 
renormalization group running of supersymmetric couplings  for the example model 
defined by eqs.~(\ref{eq:benchgAogB})-(\ref{eq:benchSMyuk}) and (\ref{eq:bencha})-(\ref{eq:benchMB}),
as a function of the renormalization scale $Q$.
The inverses of the gauge couplings $\alpha_a = g_a^2/4\pi$ are shown in the left panel.
The right panel shows a variety of renormalization group trajectories for the Yukawa coupling $y = \overline y$, obtained by taking different boundary conditions at the unification scale, illustrating the strongly attractive infrared quasi-fixed point behavior, with power-law--like running for small $y$
due to the influence of large $g_{A}$ and especially $g_{B}$.}
\end{figure}

As an illustration, Figure \ref{fig:RGrunning} shows a sample 3-loop RG trajectory, starting with an 
assumption that at the low-energy threshold scale $Q = 7.5$ TeV, where they are taken to match onto
the Standard Model,
\beq
&& 
g_{B}/g_{A} \>=\> 1.5, 
\qquad
g_{3} \>=\> 0.96171,
\label{eq:benchgAogB}
\\
&& 
g_2 \>=\> 0.628645,\qquad
g_1 \>=\> 0.366436
,
\\
&& 
y \>=\> \overline y \>=\> 2.38,
\\
&& y_t \>=\> 0.783363,\qquad
y_b \>=\> 0.012305,\qquad
y_\tau \>=\> 0.010205,
\label{eq:benchSMyuk}
\eeq
with the Standard Model Yukawa couplings chosen to correspond to $\tan\beta = 10$.
In the left panel, the gauge couplings are seen to nearly unify at a scale $7.1 \times 10^{17}$ GeV, 
much closer to the reduced Planck scale than the unification scale 
found in the MSSM. The $SU(3)_B$ coupling increases
in strength in the infrared, but does not hit a pole, 
with $\alpha_{B} = 0.239$ at $Q=7.5$ TeV; this is comparable to $\alpha_S$ evaluated 
at 3.5 GeV in the Standard Model. The chosen value of $y=\overline y$ is near the 3-loop
quasi-fixed point value for the RG equation system. 
This is illustrated in the right panel of Figure \ref{fig:RGrunning},
which shows the running for a variety of different input values. Note that even if $y$ and $\overline y$
start at much lower values (say, of order $0.1$) at the apparent unification scale, 
they are efficiently driven with power-law-like running 
in the infrared to the quasi-fixed point regime due 
to the influence of the large, and slowly running, gauge couplings $g_{A}$ and especially $g_{B}$.

The beta functions for dimensionful parameters can also be obtained from the general
results in refs.~\cite{Jones:1974pg,Jones:1983vk, West:1984dg, Parkes:1984dh,
Martin:1993yx, Martin:1993zk, Yamada:1994id, Jack:1994kd, Jack:1994rk,
Jack:1996qq, Jack:1996vg, Jack:1996cn, Jack:1997pa, Jack:1997eh,
Jack:1998iy}. 
For the supersymmetric parameter $\mu_\Phi$, one has
\beq
\beta^{(1)}_{\mu_\Phi} &=& \mu_\Phi \left [2 y^2 + 2 \overline y^2 - \frac{16}{3} (g_A^2 + g_B^2) \right ],
\\
\beta^{(2)}_{\mu_\Phi} &=& \mu_\Phi \Bigl [
\frac{128}{9} g_{A}^4  
+ \frac{256}{9} g_A^2 g_B^2
+ \Bigl (-\frac{160}{9} + \frac{16}{3} [2 n_Q + n_u + n_d] \Bigr ) g_B^4
\nonumber \\ &&
+ \frac{16}{3} (g_A^2 + g_B^2) (y^2 + \overline y^2)
- 8 y^4
- 8 \overline y^4
\Bigr ]
.
\eeq
As long as $y$ and $\overline y$ are small, this provides for $\mu_\Phi$ 
to grow rapidly in the infrared,
but this running slows as $y$ and $\overline y$ approach the quasi-fixed point regime. 
This makes it plausible that if $\mu_\Phi$ has an origin similar to that of the MSSM $\mu$ parameter,
that $\mu_\Phi$ should be larger than $\mu$ at the low scale.
In any case, it is technically natural 
for it to have any value; in the following it is assumed to be of the
same order of magnitude as the soft supersymmetry breaking masses, as could follow for example
from the Kim-Nilles \cite{KimNilles} or Giudice-Masiero \cite{GiudiceMasiero} mechanisms.
For the gaugino masses:
\beq
\beta^{(1)}_{M_{A}} &=& 0
,
\\ 
\beta^{(2)}_{M_{A}} &=& g_{A}^2 
\Bigl ( 
[192 g_{A}^2  -12 y^2 - 12 \overline y^2 - 8 y_t^2 - 8 y_b^2] M_A 
+ 32 g_{B}^2 [M_A + M_B] 
+ 18 g_2^2 [M_A + M_2] 
\nonumber \\ &&
+ \frac{22}{5} g_1^2 [M_A + M_1]
+ 12 a y + 12 \overline a \overline y + 16 a_t y_t + 16 a_b y_b
\Bigr )
,
\\
\beta^{(1)}_{M_{B}} &=& \left (-12 + 4 n_Q + 2 n_d  + 2 n_u \right ) g_{B}^2 M_B
,
\\
\beta^{(2)}_{M_{B}} &=& 
g_{B}^2 
\Bigl ( 
\{ -80 + 136 (2 n_Q + n_d + n_u)/3] g_{B}^2 -12 y^2 - 12 \overline y^2 \} M_B 
+ 32 g_{A}^2 [M_A + M_B] 
\nonumber \\ &&
+ 12 n_Q g_2^2 [M_2 + M_B] 
+ \frac{4}{15} (n_Q + 2 n_d + 8 n_u) g_1^2 [M_1 + M_B]
+ 12 a y + 12 \overline a \overline y \Bigr )
,
\\
\beta^{(1)}_{M_{2}} &=& (2 + 6 n_Q + 2 n_L) g_2^2 M_2 
,
\\
\beta^{(2)}_{M_{2}} &=& g_2^2 \Bigl (
[(100 + 84 n_Q + 28 n_L) g_2^2 - 12 y_t^2 - 12 y_b^2 - 4 y_\tau^2] M_2
+ 48 g_A^2 [M_2 + M_A]
\nonumber \\ &&
+ 32 n_Q g_B^2 [M_2 + M_B]
+ \frac{2}{5} (9 + n_Q + 3 n_L) g_1^2 [M_2 + M_1]
+ 24 a_t y_t + 24 a_b y_b + 8 a_\tau y_\tau 
\Bigr )
,
\\
\beta^{(1)}_{M_{1}} &=& 
\frac{2 g_1^3}{5} 
\left (33 + n_Q + 2 n_d + 8 n_u + 3 n_L + 6 n_3 \right ) M_1
,
\\
\beta^{(2)}_{M_{1}} &=& 
\frac{g_1^2}{5} \Bigl (
\Bigl [
(\frac{796}{5} + \frac{4}{15} n_Q + \frac{32}{15} n_d + \frac{512}{15} n_u 
+ \frac{36}{5} n_L + \frac{288}{5} n_e) g_1^2
-52 y_t^2 - 28 y_b^2 - 36 y_\tau^2
\Bigr ] M_1
\phantom{xxxx}
\nonumber \\ &&
+ 176 g_{A}^2 [M_1 + M_A] 
+ \frac{32}{3} g_{B}^2 (n_Q + 2 n_d + 8 n_u) [M_1 + M_{B}]
\nonumber \\ &&
+ (54 + 6 n_Q + 18 n_L) g_2^2 [M_1 + M_2]
+ 104 a_t y_t + 56 a_b y_b + 72 a_\tau y_\tau
\Bigr )
,
\eeq
and for the soft supersymmetry breaking parameters associated with the $\Phi, \overline \Phi$
sector: 
\beq
\beta^{(1)}_a 
&=&
18 y^2  a +  8 g_A^2 (2 y M_A - a) + 8 g_B^2 (2 y M_B  - a),
\\
\beta^{(1)}_{\overline a} 
&=&
18 \overline y^2 \overline a 
+ 8 g_A^2 (2 \overline y M_A - \overline a) + 8 g_B^2 (2 \overline y M_B - \overline a)
,
\\
\beta^{(1)}_\bphi &=& 
2 \bphi \Bigl [y^2 + \overline y^2  - \frac{8}{3} (g_A^2 + g_B^2) \Bigr ]
+ 4 \muPhi \Bigl [a y + \overline a\, \overline y + \frac{8}{3} (g_A^2 M_A + g_B^2 M_B) \Bigr ] 
,
\\
\beta^{(1)}_{m^2} &=& 12 y^2 m^2 + 4 |a|^2 - \frac{32}{3} (g_A^2 |M_A|^2 + g_B^2 |M_B|^2)
,
\phantom{xx}
\\
\beta^{(1)}_{\overline m^2} &=& 
12 \overline y^2 \overline m^2 + 4 |\overline a|^2 - \frac{32}{3} (g_A^2 |M_A|^2 + g_B^2 |M_B|^2)
.
\eeq
A consequence of these results is that if the gaugino masses are taken to be positive and large, then
$a/y$, $\overline a/\overline y$, and $b/\mu_\Phi$ tend to run to negative values 
in the conventions used here.

\begin{figure}[t]
\begin{center}
\begin{minipage}[]{0.495\linewidth}
\includegraphics[width=\linewidth,angle=0]{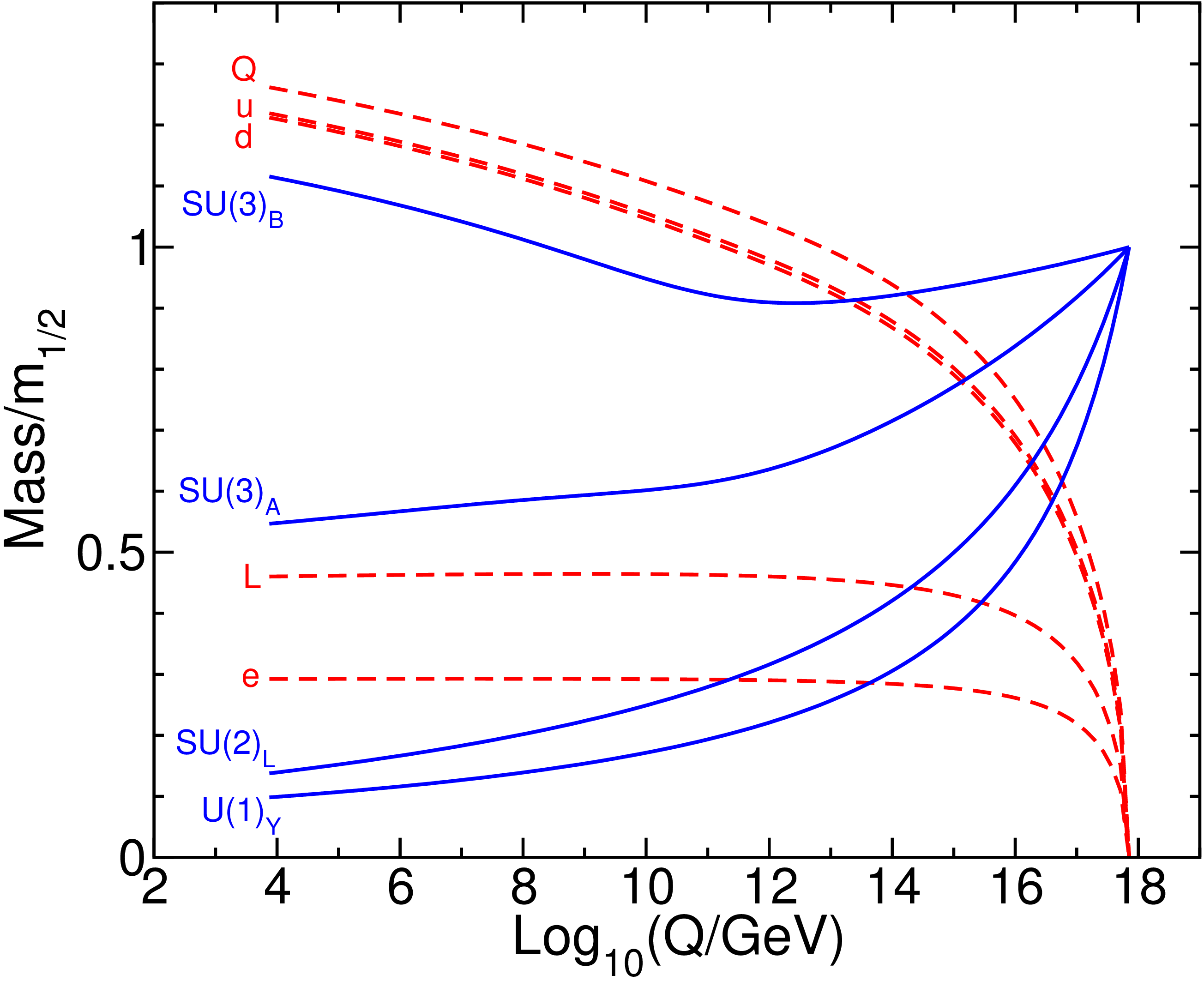}
\end{minipage}
\hspace{0.04\linewidth}
\begin{minipage}[]{0.45\linewidth}
\caption{\label{fig:RGrunningsoft}
Renormalization group running gaugino masses (solid lines) and MSSM squark and slepton masses (dashed lines),
as a function of the renormalization scale $Q$, for the example model 
defined by eqs.~(\ref{eq:benchgAogB})-(\ref{eq:benchSMyuk}) and (\ref{eq:bencha})-(\ref{eq:benchMB}),
with vanishing scalar masses and unified gaugino masses $m_{1/2}$ at the high scale.}
\end{minipage}
\end{center}
\end{figure}

The special case that will be adopted as an example 
here is the ``no-scale" limit, which presumes that at very high scales the supersymmetry breaking
is dominated by gaugino masses, with all other soft supersymmetry breaking parameters arising
from them due to renormalization group running. A nice feature of this limit is that it 
automatically provides for nearly flavor-blind first and second family squark
and slepton masses, due to the observed fact that the corresponding Yukawa couplings are small. 
As an illustration, Figure \ref{fig:RGrunningsoft} shows the running
of the gaugino masses and the MSSM squark and slepton masses as a function of the renormalization
scale $Q$, for the case of a common gaugino mass $m_{1/2}$ at the apparent unification scale.
Note that all of the
sleptons are significantly heavier than the wino and bino, in contrast to the
no-scale limit of the usual MSSM. 
This is due to the couplings $g_1$ and $g_2$ being much larger at high RG scales 
than is the case in the MSSM.
It is also worth noting that the 1-loop approximation is 
not very good, notably for $M_A$, which has an accidentally vanishing beta function at 1-loop order,
but is seen to decrease significantly in the infrared due to 2-loop and higher order effects.
The squark masses are larger than both $M_A$ and $M_B$, again in contrast to the no-scale
limit in the MSSM. Of course, these expectations could easily be modified if the high-scale boundary
conditions are different, for example due to non-universal gaugino masses. 

\subsection{Mass spectrum for an example model line\label{sec:masses}}

As an illustration of the possibilities for masses in the $SU(3)_A \times SU(3)_B$ gauge/gaugino and
$\Phi$, $\overline \Phi$ sector, consider an example model defined by the 
parameters of eqs.~(\ref{eq:benchgAogB})-(\ref{eq:benchSMyuk}) 
and the results following from renormalization group evolution 
as described in the previous subsection starting with a universal gaugino mass parameter $m_{1/2}$:
\beq
a &=& \overline a \>=\> -2.381\, m_{1/2} 
\label{eq:bencha}
,
\\
b_\phi &=& -0.6669\, m_{1/2} \mu_\Phi
,
\\
m^2 &=& \overline m^2 \>=\> (0.30806\, m_{1/2})^2
,
\\
M_A &=& 0.5467\, m_{1/2}
,
\\
M_B &=& 1.1156\, m_{1/2}
,
\label{eq:benchMB}
\eeq
where $\mu_\Phi$ is the value at the low renormalization scale. Although these values were obtained
at $Q = 7.5$ TeV, I will not commit to a particular overall mass scale for the superpartners 
or the new states in the results shown below, but instead show mass ratios normalized to
the octet vector boson mass.

The potential minimization is then found to be of the type with a real VEV given by 
eq.~(\ref{eq:realVEVxR}),
with $v = \overline v = x_R$, where $x_R$ then depends on the ratio 
$r = \mu_\Phi/m_{1/2}$. 
I vary this ratio to obtain a one-parameter model line. 
The numerical values of the dimensionless supersymmetry breaking parameters
defined in eqs.~(\ref{eq:defineA})-(\ref{eq:defineC})
are $A = -1.00034/r$, $B=-0.6669/r$, $C=0.0949/r^2$. These obey each of
the constraints in eqs.~(\ref{eq:type1req})-(\ref{eq:type1reqglobal}) for all $r$, 
and therefore yield a global minimum of the potential
at which the breaking $SU(3)_A \times SU(3)_B \rightarrow SU(3)_C$ occurs,
except for the range $0.2058 < r < 0.4611$ where it is only a local minimum due to 
a UFB solution, see eq.~(\ref{eq:nospecialUFB}). Even in that range of $r$,
the $SU(3)_C$-preserving vacuum is separated from the UFB by a barrier, making it potentially viable 
despite the UFB, if the tunneling rate is acceptably small. 
In any case, that range of $r$ will be included in the
following plots, for the sake of continuity. (Note that a slight decrease in $|B|$ would ensure that the whole range of $r$ would be a global minimum for the $SU(3)_C$ preserving vacuum.)

\begin{figure}[t]
\begin{center}
\begin{minipage}[]{0.495\linewidth}
\includegraphics[width=\linewidth,angle=0]{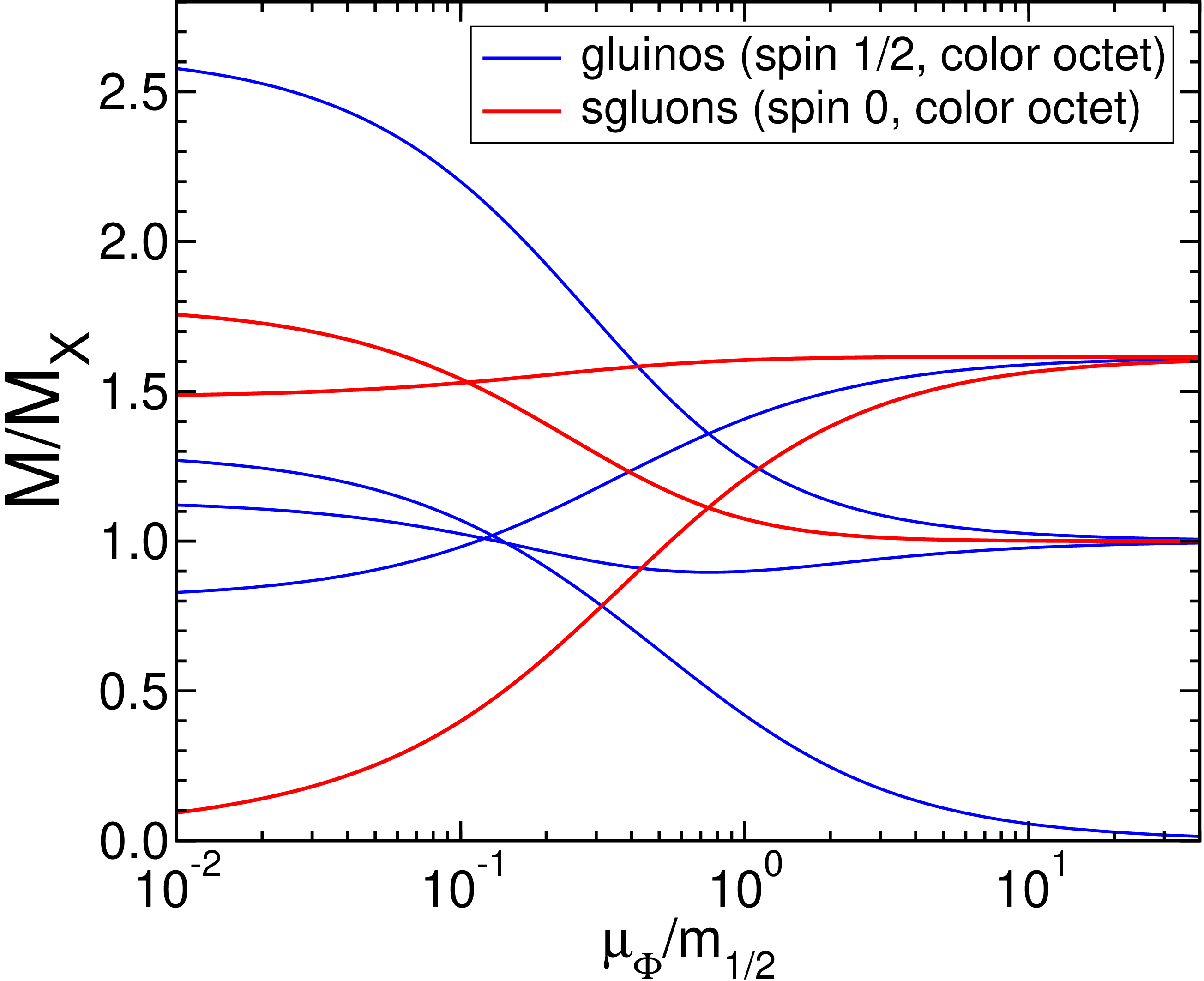}
\end{minipage}
\begin{minipage}[]{0.495\linewidth}
\includegraphics[width=\linewidth,angle=0]{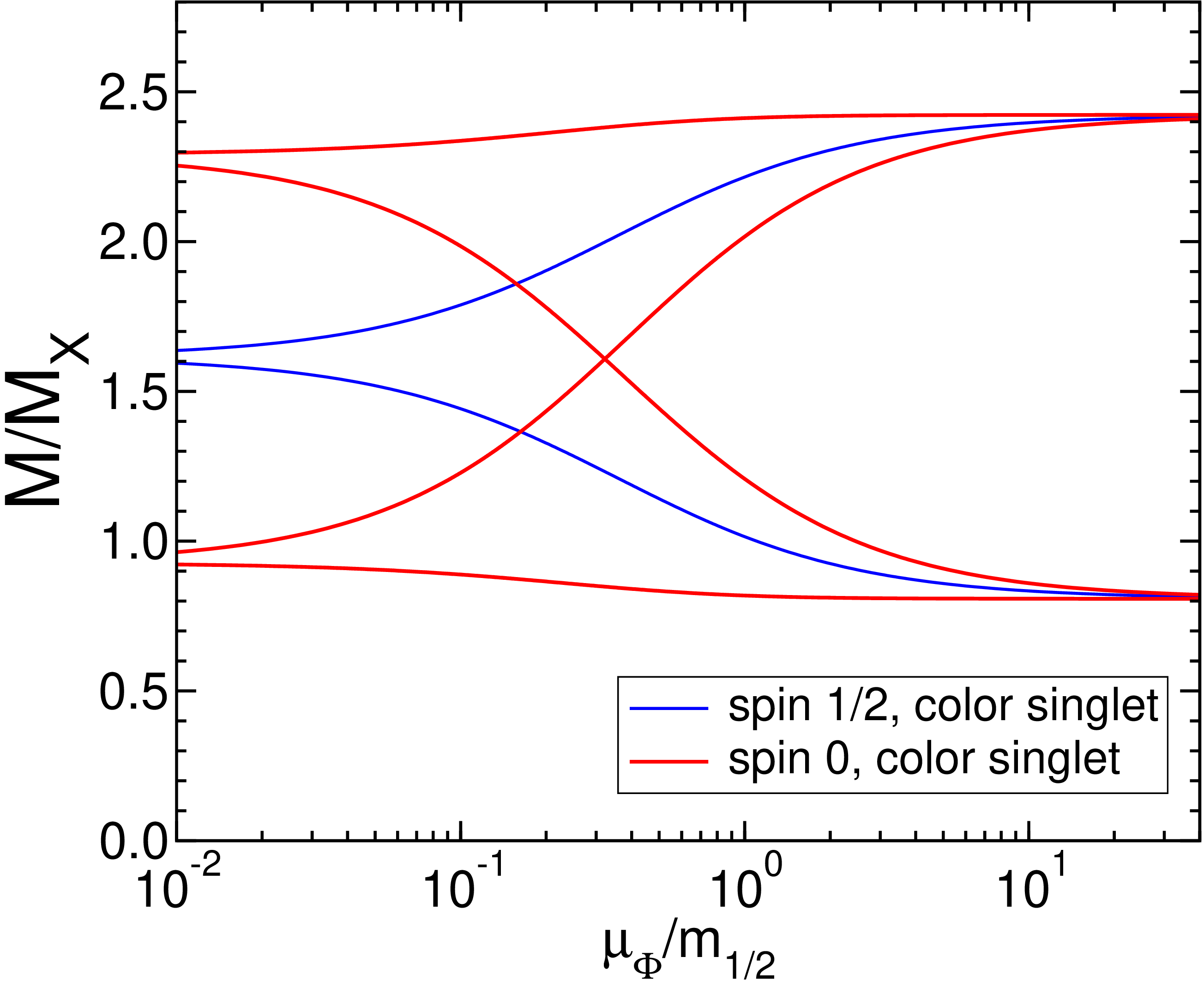}
\end{minipage}
\end{center}

\vspace{-0.5cm}

\caption{\label{fig:masses}
The ratios of fermion and scalar masses to the mass of the color octet vector boson $M_X$, as a function
of $\mu_\Phi/m_{1/2}$, for the example model line defined by 
eqs.~(\ref{eq:benchgAogB})-(\ref{eq:benchSMyuk}) and (\ref{eq:bencha})-(\ref{eq:benchMB}). 
The left panel shows the masses of the gluinos (octet fermions) and sgluons (octet scalars).
The right panel shows the masses of the color-singlet scalars and fermions from the 
$\Phi$ and $\overline \Phi$ multiplets. The right side of each plot approaches the  
supersymmetric limit, 
with masses as discussed in subsection \ref{subsec:SUSYlimit} with $R=2$.}
\end{figure}

In Figure \ref{fig:masses}, I show the masses of the four gluinos (spin-1/2 color octets),
the three sgluons (real spin-0 octets), the two singlinos (spin-1/2 color singlets) 
and the four spin-0 color singlets, all normalized to the vector (coloron) mass $M_X$.
Note that large $r = \mu_\Phi/m_{1/2}$ corresponds to the supersymmetric 
limit, in which one gluino is much lighter
than the other new states whose masses are then given by eqs.~(\ref{eq:M2vector}), 
and (\ref{eq:SUSYminVEVs})-(\ref{eq:defineRSUSY}), with $R=2$ in the present case,
which leads to $M_{\rm singlets} = \mu_\Phi$ and $3 \mu_\Phi$ and 
$M_{\rm octets} = 2 \mu_\Phi$, where $\mu_\Phi = y M_X/\sqrt{2(g_A^2 + g_B^2)} \approx 0.808 M_X$. 
In the opposite limit of small $\mu_\Phi/m_{1/2}$, the lightest of the new particles is the pseudo-scalar
sgluon.
More generally, everywhere along the model line
there is always at least one gluino state and one sgluon state and one singlet scalar
with mass below or close to
the octet vector boson mass. 

\begin{figure}[t]
\begin{center}
\begin{minipage}[]{0.495\linewidth}
\includegraphics[width=\linewidth,angle=0]{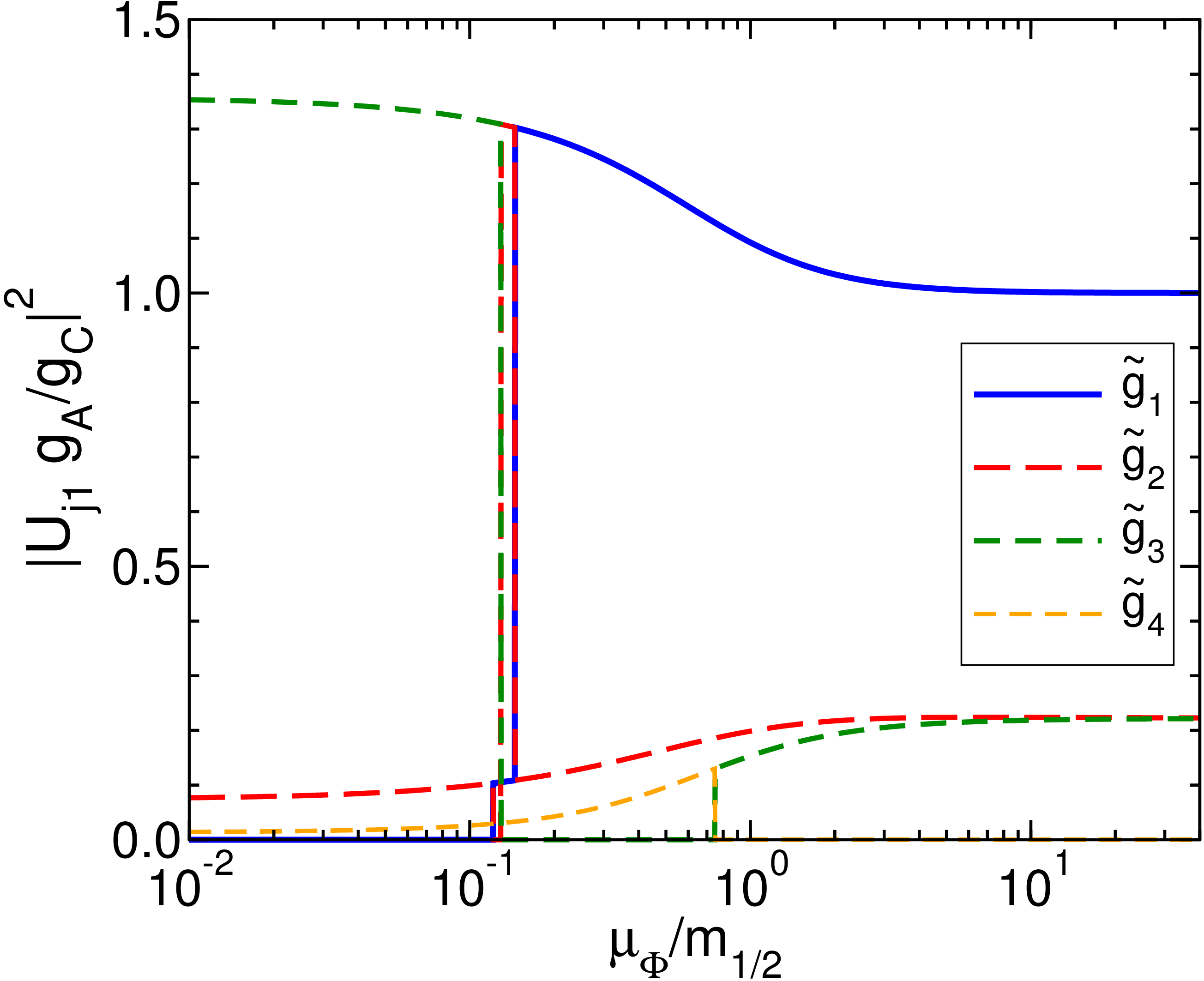}
\end{minipage}
\hspace{0.04\linewidth}
\begin{minipage}[]{0.45\linewidth}
\caption{\label{fig:gluinocouplings}
The ratio  of the squared magnitude of the coupling 
of each gluino mass eigenstate $\tilde g_j$ to quark/squark pairs to
the corresponding MSSM coupling, $|U_{j1} g_A/g_3|^2$,
 as a function of $\mu_\Phi/m_{1/2}$, for the example model line 
defined by eqs.~(\ref{eq:benchgAogB})-(\ref{eq:benchSMyuk}) 
and (\ref{eq:bencha})-(\ref{eq:benchMB}) in the text.
The right side of the plot approaches the 
supersymmetric limit.
There is always a gluino mass eigenstate with coupling to quark/squark pairs
at least as large as in the MSSM,
with the ratio of couplings for the lightest gluino approaching 1 in the supersymmetric limit.}
\end{minipage}
\end{center}
\end{figure}

In Figure \ref{fig:gluinocouplings}, for each of the four
gluino mass eigenstates ($\tilde g_j$, $j=1,2,3,4$, in increasing order of mass) 
I show the square of
the ratio of the coupling to MSSM squark/quark pairs to the corresponding coupling
that occurs in the MSSM. This is given by $|U_{j1} g_A/g_3|^2$,
in terms of the unitary matrix $U$ defined in eq.~(\ref{eq:defineUmatrix})
and the gauge couplings $g_A$ and $g_3$, governed by eq.~(\ref{eq:gCfromgAgB}). 
The result is that there is always a gluino mass eigenstate with coupling to quark/squark pairs at least 
as large as in the MSSM,
with the ratio of couplings for the lightest gluino approaching 1 in the supersymmetric limit.
However, if $\mu_\Phi/m_{1/2}$ is small, then 
the lightest gluino mass eigenstate is not MSSM-gaugino-like
and has essentially no tree-level coupling to quark-squark pairs. The
second lightest gluino state in that regime does couple to quark-squark states, 
but with a strong suppression.
The MSSM-gluino-like
state that has enhanced
couplings to quark/squark pairs can be up to about 1.6 times heavier than the lightest 
gluino state, and 1.3 times heavier than the $X$ vector boson. 
Also, in that case of small $\mu_\Phi/m_{1/2}$, the lightest new state by far is one of the sgluons;
it is possible that this would be the first new particle discovered. 
In contrast, along this model line, none of the singlinos and singlet scalars
are ever much lighter than the massive vector boson.
For all values of the ratio $\mu_\Phi/m_{1/2}$, a gluino or a sgluon is the lightest of
the non-MSSM states. 
Of course, the above results hold for a very specific set of assumptions about the RG boundary
conditions and vectorlike supermultiplet content, but I have checked that 
they are qualitatively typical at least for a (certainly non-exhaustive) 
variety of modifications of the above assumptions.

\section{Comments on collider phenomenology\label{sec:pheno}}
\setcounter{equation}{0}
\setcounter{figure}{0}
\setcounter{table}{0}
\setcounter{footnote}{1}

\baselineskip=14.99pt

The collider phenomenology of colorons, Dirac and mixed Majorana/Dirac gluinos, and sgluons has already been the subject of many papers, see refs.~\cite{Hill:1993hs,Chivukula:1996yr,
Simmons:1996fz,Bai:2010dj,Bai:2018jsr,Bai:2018wnt,Han:2010rf,Chivukula:2014pma,
Chivukula:2013xka,Chivukula:2014rka,Chen:2014haa,Chivukula:2015kua}, and 
\cite{Choi:2008pi,Kribs:2012gx,Kribs:2013eua,Kribs:2013oda,diCortona:2016fsn,Chalons:2018gez,Diessner:2019bwv}, 
and 
\cite{Plehn:2008ae,Choi:2008ub,GoncalvesNetto:2012nt,Calvet:2012rk,Beck:2015cga,Darme:2018dvz},
respectively. 
A detailed discussion of the LHC phenomenology is beyond the scope of the present paper, 
but a few brief comments are in order, with emphasis on qualitative issues where the model described 
above differs from the situation encountered in previous studies based on pure Dirac gluinos from 
supersoft and and hybrid models with an $N=2$ gauge sector. 
In this section, the particle mass eigenstates beyond those of the
MSSM\footnote{One of the $\tilde g_j$ corresponds to the MSSM gluino.
The vectorlike quarks and leptons introduced for their renormalization group running 
contributions in subsection \ref{sec:RGEs} will not be discussed; assume they are heavier.} will be denoted:
\beq
X &=& \mbox{color octet massive vector}
\\
\tilde g_j &=& \mbox{color octet gluinos, $(j=1,2,3,4)$}
\\
\tilde \chi_j &=& \mbox{color singlet fermion singlinos, $(j=1,2)$}
\\
S_j &=& \mbox{color octet real scalar sgluons, $(j=1,2,3)$}
\\
\varphi_j &=& \mbox{color singlet real scalars, $(j=1,2,3,4)$},
\eeq
where the ordering is in increasing mass.
In the example model of the previous section, the lightest of these is either $\tilde g_1$
or $S_1$. 
If $R$-parity is conserved, then the bosons have even $R$-parity and the fermions have
odd $R$-parity. 

A stringent experimental constraint comes from the fact that
the color octet vectors $X$ (colorons) have tree-level couplings to ordinary quarks, and so can 
be detected in dijet events at hadron colliders. They have a partial
width $\Gamma_X = \frac{g_A^4}{24 \pi(g_A^2 + g_B^2)} M_X$
to each flavor of quark-antiquark pair, but can also in principle have loop-induced
decays to gluon pairs.
In particular, they can be produced
singly as dijet resonances via $q \overline q \rightarrow X \rightarrow q \overline q$, resulting
in the most recent LHC bound of $M_X > 6.6$ TeV 
assuming $g_A = g_B$ 
\cite{Khachatryan:2016ecr,Sirunyan:2016iap,Aaboud:2017yvp,Aad:2019hjw,Sirunyan:2018xlo,Sirunyan:2019vgj}. 
However, in the context of the present paper the bounds will be somewhat weaker
if $g_A < g_B$, as in the example model of the previous section. 
The experimental limits also assume that the di-jet decays of $X$ dominate.
If kinematically allowed, they could also in principle decay
to squark-antisquark $\tilde q \tilde q^*$ or gluino pairs $\tilde g_j \tilde g_k$ or sgluon pairs
$S_j S_k$. 
They could even decay to $S_j \varphi_k$ (for a related study see \cite{Bai:2018wnt}) 
or $\tilde \chi \tilde g$, although these
are kinematically forbidden throughout most of the example model line of the previous section.

The gluinos $\tilde g_j$ will be pair-produced in gluon-gluon and quark-antiquark fusion, 
as is familiar from standard supersymmetry. Just as in the
MSSM, they can always decay to quark-squark final states if kinematically allowed, 
and in the alternative through virtual squarks to
$q \overline q \tilde N$ or $q \overline q' \tilde C$ where $\tilde N$ and $\tilde C$
are ordinary neutralinos and charginos. If kinematically allowed, they can also decay in a variety of
2-body modes at tree-level, to 
$\tilde g_k X$ or 
$\tilde \chi_k X$ or $\tilde g_k S$ or 
$\tilde \chi_k S$ or $\tilde g_k \varphi_l$ (if $\tilde g_k$ is a lighter gluino). 
The couplings $S \tilde g \tilde g$ and 
$S\tilde g \tilde \chi$ and $\varphi \tilde g\tilde g$ needed for the last three
decays arise from both supersymmetric gauge interactions (scalar-fermion-gaugino) and the $y,\overline y$
Yukawa couplings.  As in the MSSM, the final states of pair-produced gluino decays
will always lead to at least four jets plus missing transverse
energy signatures, sometimes with leptons from chargino or neutralino decays, and often with bottom jets
from the kinematic enhancement of lighter bottom and top squarks in the cascade decays. As noted in the previous section, one of the gluinos is likely to have an 
enhanced coupling to quark-squark pairs compared to the MSSM, unlike the case in models with 
pure or mostly Dirac gluinos. However, the gluino with enhanced couplings may not be the lightest 
gluino $\tilde g_1$. In the example model of the previous section, when $\tilde g_1$ is not 
gaugino-like and has essentially no couplings to quark-squark pairs, 
it is accompanied by a much lighter sgluon.

\begin{figure}[t]
\begin{center}
\includegraphics[width=0.99\linewidth,angle=0]{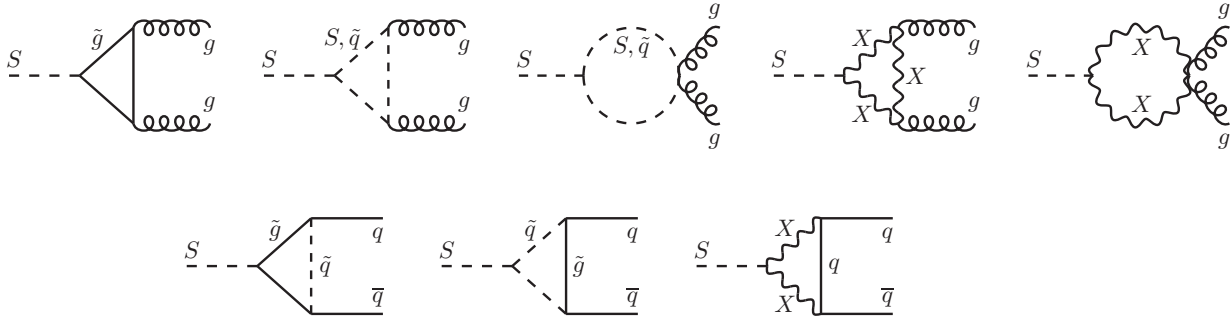}
\end{center}
\caption{\label{fig:sgluongg}
Feynman diagrams leading to effective sgluon-gluon-gluon (top row) and sgluon-quark-antiquark (bottom row)
couplings, which provide for single production and two-body decays of sgluons.}
\end{figure}

The sgluons $S$ can also be pair-produced in gluon-gluon and quark-antiquark fusion,
but can also be singly produced due to 1-loop effective couplings. The diagrams leading to
an effective $Sgg$ vertex are shown in Figure \ref{fig:sgluongg}.
Here, I note a  difference compared to the 
sgluon models previously analyzed in refs.~\cite{Plehn:2008ae,Choi:2008ub}. In those cases,
the gluino-loop contribution to the $Sgg$ vertex 
(and, more generally, the 1-loop gluino-induced 
effective couplings of $S$ to any number of gluons) was found to vanish, 
because the $S_j\tilde g_k \tilde g_l$
vertex in the 1-loop diagram was proportional to $f^{abc}$, which then requires $k\not= l$, 
causing the effective
$Sgg$ coupling to vanish, since the gluon couples only diagonally to gluino mass
eigenstates due to the unbroken QCD gauge invariance. 
However, in the model considered in the present paper,
there are also $S_j\tilde g_k \tilde g_k$ couplings proportional to the symmetric factor 
$d^{abc}$, both from the gauge couplings
$g_A$ and $g_B$ as can be seen from eq.~(\ref{eq:LSFFgauge}) and from the Yukawa couplings $y$
and $\overline y$ as seen in eq.~(\ref{eq:LSFFyukawa}). This coupling does not vanish
when inserted in the first of the loop diagrams in Figure \ref{fig:sgluongg}, although
there is a gluino mixing factor suppression. 
There is also a contribution from sgluons in the loop, in addition
to the ordinary squarks, as seen in the second and third diagrams of Figure~\ref{fig:sgluongg}.
Another difference from the models analyzed in refs.~\cite{Plehn:2008ae,Choi:2008ub} is that
massive vector loops can contribute to the effective $Sgg$ vertex, as shown in the last two 
diagrams in the first row of eq.~\ref{fig:sgluongg}. The $SXX$ vertex appearing here 
is proportional to $d^{abc}$;
it vanishes for the pseudo-scalar sgluon if CP is conserved.
These effects mean that the loop-induced $Sgg$ vertices can be significant,
and single production of $S$  due to gluon fusion can be larger than considered previously.

There are also loop-induced contributions to the
$Sq\overline q$ vertex, as shown in the second row of Figure~\ref{fig:sgluongg}, 
although this effective coupling is helicity-suppressed by the corresponding quark mass, as in
\cite{Plehn:2008ae,Choi:2008ub}. 
At tree-level the sgluons could also decay to MSSM squark-antisquark pairs if kinematically allowed, 
through the coupling inherited from the $D$-term contribution to the scalar potential.
Other two-body decays that can occur at tree-level, if kinematically allowed,
are $\tilde g \tilde g$, and $\tilde \chi\tilde g $, and $XS$, and $X\varphi$, and $\varphi S$, and $SS$.

The lightest sgluon mass eigenstate can therefore 
be produced in gluon fusion and decay to $gg$ (or to $t\overline t$), leading to
a dijet signature for which LHC searches \cite{Khachatryan:2016ecr,Sirunyan:2016iap,Sirunyan:2018xlo,Sirunyan:2019vgj,Aaboud:2017yvp,Aad:2019hjw} 
exist. However, signal/background interference
effects can be very large \cite{Martin:2016bgw} for heavy scalar di-gluon resonances, so that 
if the di-gluon production and decay dominate,
the resonance may manifest as a dip/peak or step-function invariant mass distribution
rather than a pure resonance peak. These interference effects have not been 
included in the experimental limits, 
which could be quite significantly modified if they were taken into account.

The singlet scalars $\varphi$ can likewise have loop-induced couplings to gluon-gluon and quark-antiquark.
They can therefore also be singly produced at the LHC, and would decay to jet pairs, where the
same comments just made about dijet searches apply. If kinematically
allowed, they could also decay to $\tilde g \tilde g$, $\tilde \chi \tilde \chi$, $\varphi\varphi$, $SS$, 
or $X S$ final states.

The singlino fermions $\tilde \chi$ from the $\Phi, \overline \Phi$ supermultiplets are a
new feature of the model considered here. They have tree-level 
couplings  $\tilde \chi \tilde g S$ and $\tilde \chi \tilde g X$ (proportional to gauge couplings) 
and $\tilde \chi \tilde \chi \varphi$ (proportional to Yukawa couplings $y$ and $\overline y$),
which allows for them to decay to other odd $R$-parity final states, with decay chains that will eventually terminate in the MSSM lightest
supersymmetric particle. They can always decay in this way, through off-shell intermediate 
states if necessary, so they are not stable
unless $\tilde \chi_1$ is the lightest supersymmetric particle.
However, they cannot be singly produced due to their $R$-parity, 
and cannot even be pair-produced at tree-level at colliders 
due to the lack of couplings to gluons or quarks. 
Therefore it seems unlikely that they could be part of a discovery, unless through the
cascade decays of the other states mentioned above. This also seems quite unlikely due to kinematics, 
at least for the mass spectra along the sample model line considered in the previous section.

\vspace{-15pt}

\section{Outlook\label{sec:outlook}}
\setcounter{equation}{0}
\setcounter{figure}{0}
\setcounter{table}{0}
\setcounter{footnote}{1}

If supersymmetric particles exist at a multi-TeV scale, as suggested by the tension between the big hierarchy problem
and the 125 GeV Higgs scalar boson mass, it is sensible to consider extensions of the minimal 
supersymmetric framework, including even radical ones that would not be viable at lower mass scales. 
In this paper, I have considered the possibility that the color gauge group
is extended to $SU(3) \times SU(3)$. This symmetry breaking pattern was shown to be easy to attain 
in the case of the most general renormalizable and softly broken potential that can be
constructed using the minimal field content with the necessary order parameters. Indeed, other possible 
remnant groups of the symmetry breaking were found to be highly disfavored. The model predicts several 
new color octets of spin $1$, $1/2$, and $0$, and new spin $1/2$ and spin $0$ singlets, all with masses
that are presumably at multi-TeV scales. In an example model framework motivated by 
gauge coupling unification with an infrared quasi-fixed point for the Yukawa couplings, 
gaugino-mass-dominated supersymmetry breaking leads to weakly interacting superpartners
that are relatively light, and still could be discovered at the LHC.
The phenomenology of these models at future colliders (including a high-energy LHC) will involve multiple
gluino and sgluon states, in addition to a coloron vector boson, all of which could be lighter than the
ordinary squarks. The lightest gluino could have either enhanced or highly suppressed couplings 
to quarks and squarks. The phenomenology can differ from that found in 
previous studies of Dirac and mixed gluinos and
sgluons that occur in supersoft and models with an $N=2$ gauge sector.
Although not explored here, it should also be possible to realize the same gauge symmetry breaking pattern
by introducing new singlet or octet chiral superfields. 
It is also possible to 
enlarge the gauge group that breaks down to $SU(3)_C$ in various ways. 

\noindent {\it Acknowledgments:} I am grateful to Bogdan Dobrescu for helpful conversations.
This work was supported in part by the National 
Science Foundation grant number PHY-1719273.



\begin{thebibliography}{90}
\baselineskip=15.2pt

\bibitem{Pati:1975ze} 
  J.~C.~Pati and A.~Salam,
  ``Mirror Fermions, J/psi Particles, Kolar Mine Events and Neutrino Anomaly,''
  Phys.\ Lett.\  {\bf 58B}, 333 (1975).
  doi:10.1016/0370-2693(75)90667-X

\bibitem{Preskill:1980mz} 
  J.~Preskill,
  ``Subgroup Alignment in Hypercolor Theories,''
  Nucl.\ Phys.\ B {\bf 177}, 21 (1981).
  doi:10.1016/0550-3213(81)90265-0

\bibitem{Hall:1985wz} 
  L.~J.~Hall and A.~E.~Nelson,
  ``Heavy Gluons and Monojets,''
  Phys.\ Lett.\  {\bf 153B}, 430 (1985).
  doi:10.1016/0370-2693(85)90487-3

\bibitem{Frampton:1987dn} 
  P.~H.~Frampton and S.~L.~Glashow,
  ``Chiral Color: An Alternative to the Standard Model,''
  Phys.\ Lett.\ B {\bf 190}, 157 (1987).
  doi:10.1016/0370-2693(87)90859-8

\bibitem{Frampton:1987ut} 
  P.~H.~Frampton and S.~L.~Glashow,
  ``Unifiable Chiral Color With Natural GIM Mechanism,''
  Phys.\ Rev.\ Lett.\  {\bf 58}, 2168 (1987).
  doi:10.1103/PhysRevLett.58.2168

\bibitem{Bagger:1987fz} 
  J.~Bagger, C.~Schmidt and S.~King,
  ``Axigluon Production in Hadronic Collisions,''
  Phys.\ Rev.\ D {\bf 37}, 1188 (1988).
  doi:10.1103/PhysRevD.37.1188
  
\bibitem{Hill:1991at} 
  C.~T.~Hill,
  ``Topcolor: Top quark condensation in a gauge extension of the standard model,''
  Phys.\ Lett.\ B {\bf 266}, 419 (1991).
  doi:10.1016/0370-2693(91)91061-Y

\bibitem{Martin:1992aq} 
  S.~P.~Martin,
  ``Renormalizable top quark condensate models,''
  Phys.\ Rev.\ D {\bf 45}, 4283 (1992).
  doi:10.1103/PhysRevD.45.4283,
  ``A Tumbling top quark condensate model,''
  Phys.\ Rev.\ D {\bf 46}, 2197 (1992)
  [hep-ph/9204204].

\bibitem{Hill:1993hs} 
  C.~T.~Hill and S.~J.~Parke,
  ``Top production: Sensitivity to new physics,''
  Phys.\ Rev.\ D {\bf 49}, 4454 (1994)
  [hep-ph/9312324].

\bibitem{Dobrescu:1997nm} 
  B.~A.~Dobrescu and C.~T.~Hill,
  ``Electroweak symmetry breaking via top condensation seesaw,''
  Phys.\ Rev.\ Lett.\  {\bf 81}, 2634 (1998)
  [hep-ph/9712319].
  
\bibitem{Chivukula:1998wd} 
  R.~S.~Chivukula, B.~A.~Dobrescu, H.~Georgi and C.~T.~Hill,
  ``Top Quark Seesaw Theory of Electroweak Symmetry Breaking,''
  Phys.\ Rev.\ D {\bf 59}, 075003 (1999)
  [hep-ph/9809470].
  
\bibitem{Chivukula:1996yr} 
  R.~S.~Chivukula, A.~G.~Cohen and E.~H.~Simmons,
  ``New strong interactions at the Tevatron?,''
  Phys.\ Lett.\ B {\bf 380}, 92 (1996)
  [hep-ph/9603311].

\bibitem{Simmons:1996fz} 
  E.~H.~Simmons,
  ``Coloron phenomenology,''
  Phys.\ Rev.\ D {\bf 55}, 1678 (1997)
  [hep-ph/9608269].

\bibitem{Bai:2010dj} 
  Y.~Bai and B.~A.~Dobrescu,
  ``Heavy Octets and Tevatron Signals with Three or Four b Jets,''
  JHEP {\bf 1107}, 100 (2011)
  [arXiv:1012.5814 [hep-ph]].
  
\bibitem{Chivukula:2013xka} 
  R.~S.~Chivukula, A.~Farzinnia, J.~Ren and E.~H.~Simmons,
  ``Constraints on the Scalar Sector of the Renormalizable Coloron Model,''
  Phys.\ Rev.\ D {\bf 88}, no. 7, 075020 (2013)
  Erratum: [Phys.\ Rev.\ D {\bf 89}, no. 5, 059905 (2014)]
  [arXiv:1307.1064 [hep-ph]].

\bibitem{Chivukula:2014rka} 
  R.~S.~Chivukula, E.~H.~Simmons, A.~Farzinnia and J.~Ren,
  ``LHC Constraints on a Higgs boson Partner from an Extended Color Sector,''
  Phys.\ Rev.\ D {\bf 90}, no. 1, 015013 (2014)
  [arXiv:1404.6590 [hep-ph]].

\bibitem{Chen:2014haa} 
  C.~Y.~Chen, A.~Freitas, T.~Han and K.~S.~M.~Lee,
  ``Heavy Color-Octet Particles at the LHC,''
  JHEP {\bf 1505}, 135 (2015)
  [arXiv:1410.8113 [hep-ph]].

\bibitem{Chivukula:2015kua} 
  R.~S.~Chivukula, A.~Farzinnia and E.~H.~Simmons,
  ``Vacuum Stability and Triviality Analyses of the Renormalizable Coloron Model,''
  Phys.\ Rev.\ D {\bf 92}, no. 5, 055002 (2015)
  [arXiv:1504.03012 [hep-ph]].

\bibitem{Bai:2017zhj} 
  Y.~Bai and B.~A.~Dobrescu,
  ``Minimal $SU(3)\times SU(3)$ Symmetry Breaking Patterns,''
  Phys.\ Rev.\ D {\bf 97}, no. 5, 055024 (2018)
  [arXiv:1710.01456 [hep-ph]].

\bibitem{Agrawal:2017ksf} 
  P.~Agrawal and K.~Howe,
  ``Factoring the Strong CP Problem,''
  JHEP {\bf 1812}, 029 (2018)
  [arXiv:1710.04213 [hep-ph]].

\bibitem{Bai:2018jsr} 
  Y.~Bai and B.~A.~Dobrescu,
  ``Collider Tests of the Renormalizable Coloron Model,''
  JHEP {\bf 1804}, 114 (2018)
  [arXiv:1802.03005 [hep-ph]].

\bibitem{Bai:2018wnt} 
  Y.~Bai, S.~Lu and Q.~F.~Xiang,
  ``Hexapod Coloron at the LHC,''
  JHEP {\bf 1808}, 200 (2018)
  [arXiv:1805.09815 [hep-ph]].

\bibitem{Fayet:1978qc} 
  P.~Fayet,
  ``Massive Gluinos,''
  Phys.\ Lett.\ B {\bf 78}, 417 (1978).

\bibitem{Polchinski:1982an} 
  J.~Polchinski and L.~Susskind,
  ``Breaking of Supersymmetry at Intermediate-Energy,''
  Phys.\ Rev.\ D {\bf 26}, 3661 (1982).

\bibitem{Hall:1990hq} 
  L.~J.~Hall and L.~Randall,
  ``U(1)-R symmetric supersymmetry,''
  Nucl.\ Phys.\ B {\bf 352}, 289 (1991).

\bibitem{Jack:1999ud} 
  I.~Jack and D.~R.~T.~Jones,
  ``Nonstandard soft supersymmetry breaking,''
  Phys.\ Lett.\ B {\bf 457}, 101 (1999)
  [hep-ph/9903365].

\bibitem{Jack:1999fa} 
  I.~Jack and D.~R.~T.~Jones,
  ``Quasiinfrared fixed points and renormalization group invariant trajectories for nonholomorphic soft supersymmetry breaking,''
  Phys.\ Rev.\ D {\bf 61}, 095002 (2000)
  [hep-ph/9909570].

\bibitem{supersoft} 
  P.~J.~Fox, A.~E.~Nelson and N.~Weiner,
  ``Dirac gaugino masses and supersoft supersymmetry breaking,''
  JHEP {\bf 0208}, 035 (2002)
  [hep-ph/0206096].

\bibitem{Chacko:2004mi} 
  Z.~Chacko, P.~J.~Fox and H.~Murayama,
  ``Localized supersoft supersymmetry breaking,''
  Nucl.\ Phys.\ B {\bf 706}, 53 (2005)
  [hep-ph/0406142].

\bibitem{Antoniadis:2005em} 
  I.~Antoniadis, A.~Delgado, K.~Benakli, M.~Quiros and M.~Tuckmantel,
  ``Splitting extended supersymmetry,''
  Phys.\ Lett.\ B {\bf 634}, 302 (2006)
  [hep-ph/0507192].

\bibitem{Kribs:2007ac} 
  G.~D.~Kribs, E.~Poppitz and N.~Weiner,
  ``Flavor in supersymmetry with an extended R-symmetry,''
  Phys.\ Rev.\ D {\bf 78}, 055010 (2008)
  [0712.2039].

\bibitem{Amigo:2008rc} 
  S.~D.~L.~Amigo, A.~E.~Blechman, P.~J.~Fox and E.~Poppitz,
  ``R-symmetric gauge mediation,''
  JHEP {\bf 0901}, 018 (2009)
  [0809.1112].

\bibitem{Benakli:2008pg} 
  K.~Benakli and M.~D.~Goodsell,
  ``Dirac Gauginos in General Gauge Mediation,''
  Nucl.\ Phys.\ B {\bf 816}, 185 (2009)
  [0811.4409],
%
  ``Dirac Gauginos, Gauge Mediation and Unification,''
  Nucl.\ Phys.\ B {\bf 840}, 1 (2010)
  [1003.4957].

\bibitem{Carpenter:2010as} 
  L.~M.~Carpenter,
  ``Dirac Gauginos, Negative Supertraces and Gauge Mediation,''
  JHEP {\bf 1209}, 102 (2012)
  [1007.0017].

\bibitem{Kribs:2010md} 
  G.~D.~Kribs, T.~Okui and T.~S.~Roy,
  ``Viable Gravity-Mediated Supersymmetry Breaking,''
  Phys.\ Rev.\ D {\bf 82}, 115010 (2010)
  [1008.1798].

\bibitem{Abel:2011dc} 
  S.~Abel and M.~Goodsell,
  ``Easy Dirac Gauginos,''
  JHEP {\bf 1106}, 064 (2011)
  [1102.0014].

\bibitem{Fok:2012fb} 
  R.~Fok, G.~D.~Kribs, A.~Martin and Y.~Tsai,
  ``Electroweak Baryogenesis in R-symmetric Supersymmetry,''
  Phys.\ Rev.\ D {\bf 87}, no. 5, 055018 (2013)
  [1208.2784].

\bibitem{Csaki:2013fla} 
  C.~Csaki, J.~Goodman, R.~Pavesi and Y.~Shirman,
  ``The $m_D-b_M$ problem of Dirac gauginos and its solutions,''
  Phys.\ Rev.\ D {\bf 89}, no. 5, 055005 (2014)
  [1310.4504].

\bibitem{Dudas:2013gga} 
  E.~Dudas, M.~Goodsell, L.~Heurtier and P.~Tziveloglou,
  ``Flavour models with Dirac and fake gluinos,''
  Nucl.\ Phys.\ B {\bf 884}, 632 (2014)
  [1312.2011].

\bibitem{Benakli:2014cia} 
  K.~Benakli, M.~Goodsell, F.~Staub and W.~Porod,
  ``Constrained minimal Dirac gaugino supersymmetric standard model,''
  Phys.\ Rev.\ D {\bf 90}, no. 4, 045017 (2014)
  [1403.5122].

\bibitem{Nelson:2015cea} 
A.~E.~Nelson and T.~S.~Roy,
  ``New Supersoft Supersymmetry Breaking Operators and a Solution to the $\mu$ Problem,''
  Phys.\ Rev.\ Lett.\  {\bf 114}, 201802 (2015)
  [arXiv:1501.03251 [hep-ph]].
  
\bibitem{Alves:2015kia} 
D.~S.~M.~Alves, J.~Galloway, M.~McCullough and N.~Weiner,
  ``Goldstone Gauginos,''
  Phys.\ Rev.\ Lett.\  {\bf 115}, no. 16, 161801 (2015)
  [arXiv:1502.03819 [hep-ph]].
  ``Models of Goldstone Gauginos,''
  Phys.\ Rev.\ D {\bf 93}, no. 7, 075021 (2016)
  [arXiv:1502.05055 [hep-ph]].

\bibitem{Martin:2015eca}
  S.~P.~Martin,
  ``Nonstandard Supersymmetry Breaking and Dirac Gaugino Masses without Supersoftness"
  Phys.\ Rev.\ D {\bf 92}, no. 3, 035004 (2015)
  [arXiv:1506.02105 [hep-ph]].

\bibitem{Chakraborty:2018izc}
  S.~Chakraborty, A.~Martin and T.~S.~Roy,
  ``Charting generalized supersoft supersymmetry,''
  JHEP {\bf 1805}, 176 (2018)
  [arXiv:1802.03411 [hep-ph]].

\bibitem{primer}
  S.~P.~Martin,
  ``A Supersymmetry primer,'' [hep-ph/9709356],
  version 7 (2016).

\bibitem{Jones:1974pg} 
  D.~R.~T.~Jones,
  ``Asymptotic Behavior of Supersymmetric Yang-Mills Theories in the Two Loop Approximation,''
  Nucl.\ Phys.\ B {\bf 87}, 127 (1975).
  doi:10.1016/0550-3213(75)90256-4

\bibitem{Jones:1983vk} 
  D.~R.~T.~Jones and L.~Mezincescu,
  ``The Beta Function in Supersymmetric {Yang-Mills} Theory,''
  Phys.\ Lett.\  {\bf 136B}, 242 (1984).
  doi:10.1016/0370-2693(84)91154-7

\bibitem{West:1984dg} 
  P.~C.~West,
  ``The Yukawa beta Function in N=1 Rigid Supersymmetric Theories,''
  Phys.\ Lett.\  {\bf 137B}, 371 (1984).
  doi:10.1016/0370-2693(84)91734-9

\bibitem{Parkes:1984dh} 
  A.~Parkes and P.~C.~West,
  ``Finiteness in Rigid Supersymmetric Theories,''
  Phys.\ Lett.\  {\bf 138B}, 99 (1984).
  doi:10.1016/0370-2693(84)91881-1

\bibitem{Martin:1993yx} 
  S.~P.~Martin and M.~T.~Vaughn,
  ``Regularization dependence of running couplings in softly broken supersymmetry,''
  Phys.\ Lett.\ B {\bf 318}, 331 (1993)
  [hep-ph/9308222].

\bibitem{Martin:1993zk} 
  S.~P.~Martin and M.~T.~Vaughn,
  ``Two loop renormalization group equations for soft supersymmetry breaking couplings,''
  Phys.\ Rev.\ D {\bf 50}, 2282 (1994)
  Erratum: [Phys.\ Rev.\ D {\bf 78}, 039903 (2008)]
  [hep-ph/9311340].

\bibitem{Yamada:1994id} 
  Y.~Yamada,
  ``Two loop renormalization group equations for soft SUSY breaking scalar interactions: Supergraph method,''
  Phys.\ Rev.\ D {\bf 50}, 3537 (1994)
  [hep-ph/9401241].

\bibitem{Jack:1994kd} 
  I.~Jack and D.~R.~T.~Jones,
  ``Soft supersymmetry breaking and finiteness,''
  Phys.\ Lett.\ B {\bf 333}, 372 (1994)
  [hep-ph/9405233].

\bibitem{Jack:1994rk} 
  I.~Jack, D.~R.~T.~Jones, S.~P.~Martin, M.~T.~Vaughn and Y.~Yamada,
  ``Decoupling of the epsilon scalar mass in softly broken supersymmetry,''
  Phys.\ Rev.\ D {\bf 50}, R5481 (1994)
  [hep-ph/9407291].

\bibitem{Jack:1996qq} 
  I.~Jack, D.~R.~T.~Jones and C.~G.~North,
  ``N=1 supersymmetry and the three loop anomalous dimension for the chiral superfield,''
  Nucl.\ Phys.\ B {\bf 473}, 308 (1996)
  [hep-ph/9603386].

\bibitem{Jack:1996vg} 
  I.~Jack, D.~R.~T.~Jones and C.~G.~North,
  ``N=1 supersymmetry and the three loop gauge Beta function,''
  Phys.\ Lett.\ B {\bf 386}, 138 (1996)
  [hep-ph/9606323].

\bibitem{Jack:1996cn} 
  I.~Jack, D.~R.~T.~Jones and C.~G.~North,
  ``Scheme dependence and the NSVZ Beta function,''
  Nucl.\ Phys.\ B {\bf 486}, 479 (1997)
  [hep-ph/9609325].
  
\bibitem{Jack:1997pa} 
  I.~Jack and D.~R.~T.~Jones,
  ``The Gaugino Beta function,''
  Phys.\ Lett.\ B {\bf 415}, 383 (1997)
  [hep-ph/9709364].

\bibitem{Jack:1997eh} 
  I.~Jack, D.~R.~T.~Jones and A.~Pickering,
  ``Renormalization invariance and the soft Beta functions,''
  Phys.\ Lett.\ B {\bf 426}, 73 (1998)
  [hep-ph/9712542].

\bibitem{Jack:1998iy} 
  I.~Jack, D.~R.~T.~Jones and A.~Pickering,
  ``The soft scalar mass beta function,''
  Phys.\ Lett.\ B {\bf 432}, 114 (1998)
  [hep-ph/9803405].

\bibitem{Caswell:1974gg} 
  W.~E.~Caswell,
  ``Asymptotic Behavior of Nonabelian Gauge Theories to Two Loop Order,''
  Phys.\ Rev.\ Lett.\  {\bf 33}, 244 (1974).
  doi:10.1103/PhysRevLett.33.244

\bibitem{Banks:1981nn} 
  T.~Banks and A.~Zaks,
  ``On the Phase Structure of Vector-Like Gauge Theories with Massless Fermions,''
  Nucl.\ Phys.\ B {\bf 196}, 189 (1982).
  doi:10.1016/0550-3213(82)90035-9

\bibitem{KimNilles} 
  J.~E.~Kim and H.~P.~Nilles,
  ``The mu Problem and the Strong CP Problem,''
  Phys.\ Lett.\  {\bf 138B}, 150 (1984).
  doi:10.1016/0370-2693(84)91890-2

\bibitem{GiudiceMasiero} 
  G.~F.~Giudice and A.~Masiero,
  ``A Natural Solution to the mu Problem in Supergravity Theories,''
  Phys.\ Lett.\ B {\bf 206}, 480 (1988).
  doi:10.1016/0370-2693(88)91613-9


\bibitem{Han:2010rf} 
  T.~Han, I.~Lewis and Z.~Liu,
  ``Colored Resonant Signals at the LHC: Largest Rate and Simplest Topology,''
  JHEP {\bf 1012}, 085 (2010)
  [arXiv:1010.4309 [hep-ph]].
  
\bibitem{Chivukula:2014pma} 
  R.~Sekhar Chivukula, E.~H.~Simmons and N.~Vignaroli,
  ``Distinguishing dijet resonances at the LHC,''
  Phys.\ Rev.\ D {\bf 91}, no. 5, 055019 (2015)
  [arXiv:1412.3094 [hep-ph]].


\bibitem{Choi:2008pi} 
  S.~Y.~Choi, M.~Drees, A.~Freitas and P.~M.~Zerwas,
  ``Testing the Majorana Nature of Gluinos and Neutralinos,''
  Phys.\ Rev.\ D {\bf 78}, 095007 (2008)
  [0808.2410].

\bibitem{Kribs:2012gx} 
  G.~D.~Kribs and A.~Martin,
  ``Supersoft Supersymmetry is Super-Safe,''
  Phys.\ Rev.\ D {\bf 85}, 115014 (2012)
  [1203.4821].

\bibitem{Kribs:2013eua} 
  G.~D.~Kribs and N.~Raj,
  ``Mixed Gauginos Sending Mixed Messages to the LHC,''
  Phys.\ Rev.\ D {\bf 89}, no. 5, 055011 (2014)
  [1307.7197].

\bibitem{Kribs:2013oda} 
  G.~D.~Kribs and A.~Martin,
  ``Dirac Gauginos in Supersymmetry -- Suppressed Jets + MET Signals: A Snowmass Whitepaper,''
  [1308.3468].

\bibitem{diCortona:2016fsn}
  G.~Grilli di Cortona, E.~Hardy and A.~J.~Powell,
  ``Dirac vs Majorana gauginos at a 100 TeV collider,''
  JHEP {\bf 1608}, 014 (2016)
  [arXiv:1606.07090 [hep-ph]].

\bibitem{Chalons:2018gez}
  G.~Chalons, M.~D.~Goodsell, S.~Kraml, H.~Reyes-Gonz\'alez and S.~L.~Williamson,
  ``LHC limits on gluinos and squarks in the minimal Dirac gaugino model,''
  JHEP {\bf 1904}, 113 (2019)
  [arXiv:1812.09293 [hep-ph]].

\bibitem{Diessner:2019bwv}
  P.~Diessner, J.~Kalinowski, W.~Kotlarski and D.~St\"ockinger,
  ``Confronting the coloured sector of the MRSSM with LHC data,''
  JHEP {\bf 1909}, 120 (2019)
  [arXiv:1907.11641 [hep-ph]].

\bibitem{Plehn:2008ae} 
  T.~Plehn and T.~M.~P.~Tait,
  ``Seeking Sgluons,''
  J.\ Phys.\ G {\bf 36}, 075001 (2009)
  [0810.3919].

\bibitem{Choi:2008ub} 
  S.~Y.~Choi, M.~Drees, J.~Kalinowski, J.~M.~Kim, E.~Popenda and P.~M.~Zerwas,
  ``Color-Octet Scalars of N=2 Supersymmetry at the LHC,''
  Phys.\ Lett.\ B {\bf 672}, 246 (2009)
  [0812.3586].

\bibitem{GoncalvesNetto:2012nt} 
  D.~Goncalves-Netto, D.~Lopez-Val, K.~Mawatari, T.~Plehn and I.~Wigmore,
  ``Sgluon Pair Production to Next-to-Leading Order,''
  Phys.\ Rev.\ D {\bf 85}, 114024 (2012)
  [arXiv:1203.6358 [hep-ph]].

\bibitem{Calvet:2012rk} 
  S.~Calvet, B.~Fuks, P.~Gris and L.~Valery,
  ``Searching for sgluons in multitop events at a center-of-mass energy of 8 TeV,''
  JHEP {\bf 1304}, 043 (2013)
  [arXiv:1212.3360 [hep-ph]].

\bibitem{Beck:2015cga} 
  L.~Beck, F.~Blekman, D.~Dobur, B.~Fuks, J.~Keaveney and K.~Mawatari,
  ``Probing top-philic sgluons with LHC Run I data,''
  Phys.\ Lett.\ B {\bf 746}, 48 (2015)
  [arXiv:1501.07580 [hep-ph]].
  
\bibitem{Darme:2018dvz}
  L.~Darm\'e, B.~Fuks and M.~Goodsell,
  ``Cornering sgluons with four-top-quark events,''
  Phys.\ Lett.\ B {\bf 784}, 223 (2018)
  [arXiv:1805.10835 [hep-ph]].

  
\bibitem{Khachatryan:2016ecr} 
  V.~Khachatryan {\it et al.} [CMS Collaboration],
  ``Search for narrow resonances in dijet final states at $\sqrt(s)=$ 8 TeV with the novel CMS technique of data scouting,''
  Phys.\ Rev.\ Lett.\  {\bf 117}, no. 3, 031802 (2016)
  [arXiv:1604.08907 [hep-ex]].

\bibitem{Sirunyan:2016iap} 
  A.~M.~Sirunyan {\it et al.} [CMS Collaboration],
  ``Search for dijet resonances in proton?proton collisions at $\sqrt{s}$ = 13 TeV and constraints on dark matter and other models,''
  Phys.\ Lett.\ B {\bf 769}, 520 (2017)
  Erratum: [Phys.\ Lett.\ B {\bf 772}, 882 (2017)]
  [arXiv:1611.03568 [hep-ex]].

\bibitem{Aaboud:2017yvp} 
  M.~Aaboud {\it et al.} [ATLAS Collaboration],
  ``Search for new phenomena in dijet events using 37 fb$^{-1}$ of $pp$ collision data collected at $\sqrt{s}=$13 TeV with the ATLAS detector,''
  Phys.\ Rev.\ D {\bf 96}, no. 5, 052004 (2017)
  doi:10.1103/PhysRevD.96.052004
  [arXiv:1703.09127 [hep-ex]].

\bibitem{Aad:2019hjw} 
  G.~Aad {\it et al.} [ATLAS Collaboration],
  ``Search for new resonances in mass distributions of jet pairs using 139 fb$^{-1}$ of $pp$ collisions at $\sqrt{s}=13$ TeV with the ATLAS detector,''
  arXiv:1910.08447 [hep-ex].

\bibitem{Sirunyan:2018xlo} 
  A.~M.~Sirunyan {\it et al.} [CMS Collaboration],
  ``Search for narrow and broad dijet resonances in proton-proton collisions at $ \sqrt{s}=13 $ TeV and constraints on dark matter mediators and other new particles,''
  JHEP {\bf 1808}, 130 (2018)
  [arXiv:1806.00843 [hep-ex]].
  
\bibitem{Sirunyan:2019vgj} 
  A.~M.~Sirunyan {\it et al.} [CMS Collaboration],
  ``Search for high mass dijet resonances with a new background prediction method in proton-proton collisions at $\sqrt{s}=$ 13 TeV,''
  arXiv:1911.03947 [hep-ex].

\bibitem{Martin:2016bgw} 
  S.~P.~Martin,
  ``Signal-background interference for a singlet spin-0 digluon resonance at the LHC,''
  Phys.\ Rev.\ D {\bf 94}, no. 3, 035003 (2016)
  [arXiv:1606.03026 [hep-ph]].
  
\end{thebibliography}
\end{document}